\documentclass[12pt,preprint]{aastex}

\newcommand{\teff}{$T_{\rm eff}$}
\newcommand{\gtau}{$\gamma \; \mathrm{Tau}$}
\newcommand{\dtau}{$\delta \; \mathrm{Tau}$}
\newcommand{\etau}{$\epsilon \; \mathrm{Tau}$}
\newcommand{\cara}{$^{12}\mathrm{C}$}
\newcommand{\carb}{$^{13}\mathrm{C}$}
\newcommand{\nit}{$^{14}\mathrm{N}$}
\newcommand{\oxy}{$^{16}\mathrm{O}$}

\begin{document}

\title{STELLAR NUCLEOSYNTHESIS IN THE HYADES OPEN CLUSTER\altaffilmark{1}}

\altaffiltext{1}{Based on data taken with the Harlan J. Smith 2.7-m telescope 
at The McDonald Observatory of the University of Texas at Austin.}

\author{Simon C. Schuler\altaffilmark{2,4}, Jeremy R. King\altaffilmark{3}, 
AND Lih-Sin The\altaffilmark{3}}
\affil{
   \altaffiltext{2}{National Optical Astronomy Observatory, 950 North Cherry
   Avenue, Tucson, AZ, 85719; sschuler@noao.edu}
   \altaffiltext{3}{Department of Physics and Astronomy, Clemson University, 118
   Kinard Laboratory, Clemson, SC, 29634; jking2@ces.clemson.edu, 
   tlihsin@clemson.edu}
   \altaffiltext{4}{Leo Goldberg Fellow}
   }

\begin{abstract}
We report a comprehensive light element (Li, C, N, O, Na, Mg, and Al) abundance 
analysis of three solar-type main sequence (MS) dwarfs and three red giant branch 
(RGB) clump stars in the Hyades open cluster using high-resolution and high 
signal-to-noise spectroscopy.  The abundances have been derived in a self-consistent
fashion, and for each group (MS or RGB), the CNO abundances are found to be in 
excellent star-to-star agreement.  Using the dwarfs to infer the initial composition 
of the giants, the combined abundance patterns confirm that the giants have 
undergone the first dredge-up and that material processed by the CN cycle has been 
mixed to the surface layers.  The observed abundances are compared to predictions 
of a standard stellar model based on the Clemson-American University of Beirut 
(CAUB) stellar evolution code.  The model reproduces the observed evolution of the 
N and O abundances, as well as the previously derived \cara/\carb\ ratio, but it 
fails to predict by a factor of 1.5 the observed level of \cara\ depletion.  A 
similar discord appears to exist in previously reported observed and modeled C 
abundances of giants in the Galactic disk.  Random uncertainties in the mean 
abundances and uncertainties related to possible systematic errors in the Hyades 
dwarf and giant parameter scales cannot account for the discrepancy in the observed 
and modeled abundances.  Li abundances are derived to determine if non-canonical 
extra mixing, like that seen in low-mass metal-poor giants, has occurred in the 
Hyades giants.  The Li abundance of the giant \gtau\ is in good accord with the 
predicted level of surface Li dilution, but a $\sim 0.35$ dex spread in the giant 
Li abundances is found and cannot be explained by the stellar model.  Possible 
sources of the spread are discussed; however, it is apparent that the differential 
mechanism responsible for the Li dispersion must be unrelated to the uniformly low 
\cara\ abundances of the giants.  Na, Mg, and Al abundances are derived as an 
additional test of our stellar model.  All three elements are found to be 
overabundant by 0.2 -- 0.5 dex in the giants relative to the dwarfs.  Such large 
enhancements of these elements are not predicted by the stellar model, and non-LTE 
effects significantly larger (and, in some cases, of opposite sign) than those 
implied by extant literature calculations are the most likely cause.  
\end{abstract}


\section{INTRODUCTION}
\label{s:intro}
The CN cycle is the dominant energy source in the H burning cores of main sequence 
(MS) stars more massive than the Sun.  The cycle effectively converts four protons 
into a $^{4}$He nucleus using \cara, \carb, \nit, and $^{15}$N as catalysts for the
reactions.  If O is present, the ON cycle can inject \nit\ into the CN cycle at 
the expense of \oxy.  The combined CNO bi-cycle produces no net loss of CNO 
nuclei, but their relative abundances are altered due to their different lifetimes 
against proton capture.  The slowest reaction in the cycle, \nit 
($p,\gamma$)$^{15}$O, creates a bottleneck, and by the time the reactions achieve 
a steady state, the abundance of \cara\ and the related \cara/\carb\ ratio have 
been reduced, and the abundance of \nit\ has been increased.  

Prior to the onset of core He burning, the convective envelope extends from the 
stellar surface down to the inner layers, reaching depths where the CN cycle had 
been previously active \citep{1964ApJ...140.1631I}.  Known as the first dredge-up, 
the extended convective envelope chemically homogenizes the outer and inner layers 
down to an interior mass of about 0.5 M$_{\odot}$ (depending on the mass and 
metallicity of the progenitor star), and products of the core nuclear processes 
once shrouded by the star's optically thick layers are now brought to the surface 
where they can be observed.  No additional alteration of the surface composition 
is expected until the star evolves off of the red giant branch (RGB).

Many high-resolution spectroscopic studies have targeted CNO and other light 
elements such as Li, Na, Mg, and Al (the abundances of which may be affected by 
core nuclear processes and the first dredge-up) in order to verify the general 
scenario outlined above and to test the quantitative predictions of stellar 
evolution models.  Observed CNO abundance patterns, and \cara/\carb\ ratios, 
of near-solar metallicity giants in general have been found to be in good 
agreement with standard stellar evolution models 
\citep[e.g.,][]{1981ApJ...248..228L,2007AJ....133.2464L}, but there are exceptions.
\citet{1989ApJ...347..835G} found \cara/\carb\ ratios of $15 \pm 3$ for some open 
cluster giants, well below typical predicted values of about 23 -- 25
\citep[e.g.,][]{2002PASP..114..375S}.  Standard models are those in which mixing 
is done by convection only; non-canonical mixing mechanisms such as 
rotationally-induced mixing are not included.  

Standard models are also unable to reproduce the light element abundances of 
low-mass ($M \lesssim 2.5$ M$_{\odot}$) metal-poor giants brighter than the RGB 
bump.  These giants have \cara/\carb\ ratios that are generally below 10 and often 
reach near the equilibrium value of $\sim 3.5$; they also have depleted \cara\ and
enhanced \nit\ abundances relative to giants at the bump and essentially no Li 
\citep[e.g.,][]{2000A&A...354..169G,2003ApJ...585L..45S}.  The observed abundance 
patterns are believed to be the result of an extra mixing episode subsequent to 
the RGB bump, and shear instabilities and thermohaline mixing have been shown to 
be promising non-canonical mixing mechanisms that can account for the observations 
\citep[e.g.,][]{2006ApJ...641.1087D,2007A&A...467L..15C}.  

Despite overall good agreement between observations and stellar evolution models, 
it is clear that standard models are incomplete, and light element abundance 
analyses of giants with different masses and metallicities will continue to 
provide valuable empirical constraints for future theoretical efforts.  The 
chemical homogeneity of open clusters make them excellent laboratories for this 
purpose.  Open cluster dwarf abundances can be used as a proxy for the initial 
metallicity of giants in the same cluster, allowing the surface abundance 
evolution resulting from the first dredge-up to be empirically quantified.  The 
light element abundances of open cluster giants have been derived by numerous 
previous groups \citep[e.g.,][]{1977ApJ...217..508L,1981ApJ...248..228L,1985ApJ...297..233B,1989ApJ...347..835G,2000A&A...360..499T},
but none have simultaneously determined dwarf and giant abundances from the same 
data set.  \citet{2004A&A...424..951P} performed a detailed chemical analysis of 
22 stars, ranging from MS dwarfs to post-RGB clump giants, in the open cluster IC
4651.  Li, Na, Mg, and Al were included in the analysis, but CNO were not. 

We report here the first comprehensive light-element abundance analysis of 
both MS dwarfs and RGB clump giants in the Hyades open cluster carried-out in a 
self-consistent fashion.  Abundances of Li, C, N, Na, Mg, and Al are newly derived 
from a homogeneous set of high-resolution echelle spectra, and combined with O
abundances previously derived from the same spectra \citep{2006AJ....131.1057S}, 
we conduct a critical comparison between the observed light-element abundances 
and the predictions of a standard stellar evolution model tailored to the Hyades 
giants.

\section{STELLAR EVOLUTION MODEL}
\label{s:model}
We have employed the Clemson-American University of Beirut (CAUB) stellar 
evolution code \citep{2000ApJ...533..998T,2004ApJ...611..452E,2007ApJ...655.1058T},
a 1D implicit hydrodynamical Lagrangian code, to model the surface compositions of 
the Hyades giants.  In its general form, the code evolves a star by first solving 
the stellar structure equations, then by performing the nuclear burning for the 
given time step, and finally by diffusively mixing the nuclear species present in 
the convective zones.  However, the code has been improved recently so that it can 
carry out simultaneously the nuclear burning and mixing stages.  This allows the
inclusion of the Cameron-Fowler mechanism, the nucleosynthesis and mixing 
timescales for which are quite similar, and its possible effects on the Li surface 
abundance.  We find that the nucleosynthesis results of both recipes are in good 
agreement up to the end of core He burning; for example, no significant difference 
in the $^{7}$Li surface abundance before and after the first dredge-up is seen 
between the two calculations.

The thermonuclear reaction rates included in the CAUB code are from the NACRE 
collaboration \citep{1999NuPhA.656....3A} and are complemented by the NonSmoker 
compilation of \citet{2000ADNDT..75....1R}.  The nuclear data are taken from the 
study of \citet{1995NuPhA.595..409A}.  For the important reaction 
$^{12}$C($\alpha$,$\gamma$)$^{16}$O, we adopt the rate given by 
\citet{2002ApJ...567..643K}, who have reduced considerably the uncertainties in the 
relevant cross section at stellar temperatures.  The code uses an electron-positron 
equation of state (EOS) based on table interpolation of the Helmholtz Free energy 
of \citet{2000ApJS..126..501T}.  The EOS solves the Saha equation for partially 
ionized matter, and it is accurate for any degree of degeneracy and relativistic 
matter covering 10$^{-12}$ $\le \rho \le $ 10$^{15}$ g cm$^{-3}$ and 10$^3$ $\le T 
\le $ 10$^{13}$ K.  EOS\_2005 Rosseland mean opacities from Livermore 
\citep{1996ApJ...464..943I} have been used for the H-rich and exhausted H 
compositions.  For the stellar layers with temperatures lower than $\sim$6000 K, 
the Rosseland mean opacities calculated by \citet{2005ApJ...623..585F} are used.  
The code makes use of the Schwarzchild criterion of convection in determining the 
convective zones in the stellar model; neither overshooting nor semiconvection is 
included in the calculations.  A mixing length parameter $\alpha$ = 2.0 ($l = 
\alpha \; H_p$, where l is the convective scale length and $H_p$ is the local 
pressure scale height) has been adopted.  Because mass loss is expected to be minor 
in low mass stars, it is not included in the evolution of the model.

The stellar model, which is characterized by a mass of 2.5 M$_{\odot}$ and an 
initial metallicity of Z = 0.025 ([m/H] $\simeq$ +0.10\footnotemark[5]), is an 
updated version of the one used in our previous Hyades study, where the mass and 
metallicity of the model are discussed \citep{2006AJ....131.1057S}.  Throughout this 
paper we adopt the Hyades metallicity (i.e., [Fe/H]) of 
\citet[][$[\mathrm{Fe/H]} = +0.13$]{2003AJ....125.3185P}, and the metallicity of 
the stellar model was chosen to closely match this value.  The model begins at the 
zero main sequence phase with $\log T_{\mathrm{eff}} = 4.027$, $\log (L/L_{\odot}) 
= 1.632$, central temperature $T_{\mathrm{C}} = 2.28 \times 10^7 \; \mathrm{K}$, 
and central density $\rho_{\mathrm{C}} = 4.687 \times 10^1 \; \mathrm{g cm}^{-3}$.  
At this time, the H convective core is at its maximum mass, 0.436 M$_{\odot}$.  The 
model reaches the end of the core H burning phase when its central nuclear energy 
generation is at a minimum, which occurs after an evolution time of $5.121 \times 
10^8$ yr.  At this stage of the calculation, the star is characterized by a $\log 
T_{\mathrm{eff}} = 3.921$, $\log (L/L_{\odot}) = 1.824$, $T_{\mathrm{C}} = 
2.57\times 10^7 \; \mathrm{K}$, $\rho_{\mathrm{C}} = 7.678 \times 10^2 \; \mathrm{g 
cm}^{-3}$, and a central $^{4}$He mass fraction $X(^{4}\mathrm{He}) = 0.9786$.  Up until 
this point, the photospheric elemental mass fractions of the MS star have not been 
altered from their initial values.

\footnotetext[5]{Throughout this paper we use the standard bracket notation to denote
stellar abundances given relative to solar values, e.g., [m/H] = $\log 
[N$(m)/$N$(H)]$_{*} - \log [N$(m)/$N$(H)]$_{\odot}$ on a scale where $\log N$(H) = 
12.0.} 

After the end of the core H burning stage approximately $1.275 \times 10^7 \; 
\mathrm{yr}$ later, the first dredge-up starts to develop and reaches a maximum 
depth at an interior radius of $M_{\mathrm{r}} = 0.338 \; \mathrm{M_{\odot}}$, 
before the He convective core begins to develop.  During the inward progression of 
the convective envelope (which lasts only for $\sim 1.51 \times 10^7$ yr), the 
elemental surface mass fractions show some transformation due to the mixing of 
envelope material with the products of the core H burning.  After the first 
dredge-up, no additional alteration of the surface abundances is seen until the 
onset of the third dredge-up (during shell He burning).  During core He burning, 
the bottom of the convective envelope retracts outward to an interior radius of 
$M_{\mathrm{r}} = 1.480 \; \mathrm{M_{\odot}}$.  The convective He core grows up to 
a maximum of 0.242 M$_{\odot}$, with He burning in the core lasting for $\sim 2.82 
\times 10^8 \; \mathrm{yr}$.  Observationally, the Hyades giants are currently 
residing at the cluster RGB clump \citep{2001A&A...367..111D}, and are thus in 
their core He-burning stages.

\subsection{Evolution of Surface Abundances}
\label{ss:predictions}

Figure 1 shows the composition profiles after the end of core H burning, before the 
first dredge-up of our 2.5 M$_{\odot}$ stellar model; Figure 2 shows the 
composition profiles after the first dredge-up but before core He burning ignites.  
The first dredge-up lasts a very short time ($\sim 1.5 \times 10^7$ yrs) relative 
to the duration of core H ($5.12 \times 10^8$ yrs) or core He burning ($\sim 2.8 
\times 10^8$ yrs).  However, many isotopic surface abundances change dramatically 
during the dredge-up process.
\marginpar{Fig.1} \marginpar{Fig.2}

Comparing the energy generation rate of the proton-proton (p-p) chains with that of 
the CNO bi-cycle we find that a star with a mass larger than $\sim 1.3$ M$_{\odot}$ 
or having a central temperature greater than $\sim 1.7 \times 10^7$ K has its 
energy generation dominated by the CNO bi-cycle \citep{1996snai.book.....A}.  To 
understand the features of the composition profiles in Figure 1, we look at the 
properties of the p-p chains, CNO bi-cycle \citep{1968psen.book.....C}, and the 
NeNa and MgAl cycles \citep{1988ccna.book.....R}.  Examining the profiles from the 
surface inward, the temperature in the star increases, and we notice the first 
observable consequence of the p-p chains, i.e., the destruction of $^{7}$Li through 
$^{7}$Li(p,$\alpha$)$^{4}$He reaction ($T \ga 3.8 \times 10^6$ K).  Somewhat 
further inward, we encounter the location where most of the $^{10}$B has been 
destroyed through the $^{10}$B(p,$\alpha$)$^7$Li reaction ($T \ga 6.0 \times 10^6$ 
K).  During the early evolution of H burning, the $^7$Li and $^{10}$B profiles 
trail each other as their steep destruction profiles move further outward.  
Furthermore, due to their steep profiles and their proximity in the stellar 
atmosphere, the mass-thickness ratio of the $^7$Li and $^{10}$B profiles before the 
first dredge-up can be estimated by the $^7$Li and $^{10}$B abundance ratios in the 
dwarf and giant stars.  Further inward, we find another product of the p-p chains, 
$^{3}$He, that was predominantly produced through $^2$D(p,$\gamma$)$^3$He ($T \ga
7.7 \times 10^6$ K).  However, at higher temperatures ($T \ga 13.5 \times 10^6$ K), 
$^3$He is predominantly self-destructed through $^3$He($^3$He,2p)$^4$He reactions.

With respect to the CNO bi-cycle, the first effective reaction sequence at low 
temperature ($T \ga 10.2 \times 10^6$ K) is the CN cycle, 
$^{12}$C(p,$\gamma$)$^{13}$N(e$^+\nu$)$^{13}$C(p,$\gamma$)$^{14}$N(p,$\gamma$)$^{15}$O(e$^+\nu$)$^{15}$N(p,$\alpha$)$^{12}$C.
The CN cycle starts by depleting $^{12}$C and producing $^{13}$C, which is then
converted predominantly into $^{14}$N; the initial abundance of $^{15}$N is also
depleted.  The steady state abundances of the secondary nuclei $^{13}$C and 
$^{15}$N are a function of temperature, such that there is a peak of $^{13}$C and a 
valley of $^{15}$N abundances and an equilibrium ratio of $^{15}$N/$^{14}$N 
(horizontal sections of $^{14}$N and $^{15}$N curves).

Further inward where the temperatures are higher ($T \ga 13.2 \times 10^6$ K), the 
ON cycle becomes effective, resulting in the production of $^{17}$O and the 
depletion of \oxy\ and $^{18}$O through 
$^{16}$O(p,$\gamma$)$^{17}$F(e$^+\nu$)$^{17}$O and $^{18}$O(p,$\alpha$)$^{15}$N 
reactions, respectively.  Approaching the previously convective core, the 
abundances of $^{12}$C, $^{13}$C, $^{17}$O, and $^{18}$O increase, as does that of 
$^{23}$Na, indicating the action of the NeNa cycle.  At these high temperatures 
($T \ga 15 \times 10^6$ K), the $^{22}$Ne(p,$\gamma$)$^{23}$Na reaction depletes 
$^{22}$Ne and produces $^{23}$Na; the consequence of this reaction is that any 
surface abundance enhancement in $^{23}$Na should be accompanied by the concomitant 
reduction in $^{22}$Ne by approximately the same amount by number.

As noticed by \cite{1994A&A...285..915E}, inside the H convective core the CNO 
bi-cycle reaches an equilibrium state, and all O isotopes and $^{19}$F are 
depleted.  On the contrary, $^{14}$N and $^{4}$He are strongly produced together, 
accompanied by the enhancements in $^{12}$C and $^{13}$C.  In Figure 3, we show the 
ratio of the surface abundance before and after the first dredge-up.  Enhancements 
of surface abundances, ranked from the largest to the smallest, are $^{17}$O, 
$^{3}$He, $^{13}$C, $^{14}$N, $^{23}$Na, and $^4$He.  Reductions of surface 
abundances, ranked from the largest to the smallest, are $^{7}$Li, $^{10}$B, 
$^{15}$N, $^{12}$C, $^{18}$O, $^{22}$Ne, $^{16}$O, and $^{19}$F.  The ranking shows 
the relative degree of sensitivity required for dwarf and giant observations to 
detect the dredge-up process.
\marginpar{Fig.3}

\section{HIGH-RESOLUTION SPECTROSCOPY: ANALYSIS \& ABUNDANCES}
\subsection{The Spectra}
\label{ss:spectra}
High-resolution, high signal-to-noise (S/N) ratio spectra of our Hyades sample 
were obtained with the Harlan J. Smith 2.7-m telescope and the``2dcoude" 
cross-dispersed echelle spectrometer at the McDonald Observatory in 2004 
October.  The instrument configuration and spectra have been described 
previously in \citet{2006ApJ...636..432S}.  The nominal spectral resolution of 
our data is $R = \lambda / \Delta \lambda \approx 60,000$ ($\sim 2.1 \; 
\mathrm{pixels}$), with a typical S/N ratio of 150-200 for the dwarfs and 
400-600 for the giants.  Reductions of the spectra followed standard routines for 
bias subtraction, flat fielding, scattered light removal, order extraction, and 
wavelength calibration using the {\sf IRAF}\footnotemark[6] package.

\footnotetext[6]{IRAF is distributed by the National Optical Astronomy 
Observatory, which is operated by the Association of Universities for Research 
in Astronomy, Inc., under cooperative agreement with the National Science 
Foundation.}

Additional spectra of the Hyades giants were obtained with the Harlan J. Smith 
2.7-m telescope and the 2dcoude cross-dispersed echelle spectrometer on 2008
March 18.  The instrument configuration included the cs23 setting, E2 echelle 
grating, TK3 detector with 24 $\mu$m pixels, and a $1 \times 1$ pixel binning, 
resulting in a nominal resolving power of $R = 60,000$.  This configuration 
matches exactly that from our 2004 October observations except for the echelle 
grating position, which was set in the more recent observations to maximize 
throughput in the 8727 {\AA} [\ion{C}{1}] region.  Two 240 s integrations were 
taken of each $\gamma$, $\delta$, and $\epsilon$ Tau, resulting in S/N ratios of 
365, 495, and 590, respectively, in the $\lambda8727$ region.

\subsection{Stellar Parameters}
\label{ss:params}
The stars used for the current analysis are a subset of our larger Hyades 
sample for which we have high-S/N, high-resolution spectra.  Stellar parameters 
have been derived in a consistent manner for our entire Hyades dwarf sample
using published photometry for \teff, the $Y^2$ isochrones 
\citep{2003ApJS..144..259Y} for $\log g$, and the empirical relation of 
\citet{2004A&A...420..183A} for microturbulent velocity ($\xi$).  For the giants,
the stellar parameters have been collated from the literature.  The derivation 
and adoption of the stellar parameters are described in 
\citet{2006AJ....131.1057S,2006ApJ...636..432S}, where a discussion of the
related uncertainties can also be found.  The stellar parameters of the current 
sample are given in Table 1.  The adopted uncertainties in the stellar parameters 
are provided in the footnotes of Table 1; these uncertainties are the
uncertainties in the absolute stellar parameters and are most appropriate for
estimating the uncertainties in the derived dwarf and giant abundances.  We note
that the relative parameter uncertainties between the dwarfs themselves or between
the giants themselves are likely smaller.
\marginpar{Tab.1}

\subsection{CNO Abundances}
\label{ss:cnoAbunds}
The chemical abundances presented herein have been derived by means of a local
thermodynamic equilibrium (LTE) analysis using an updated version of MOOG, the 
LTE line analysis and spectrum synthesis software package (Sneden 1973; C. Sneden 
2004, private communication).  Model atmospheres used in the analyses were 
interpolated from Kurucz ATLAS9 grids generated with the convective overshoot 
approximation and are the same as those used in our previous Hyades studies 
\citep{2006AJ....131.1057S,2006ApJ...636..432S}.  The method and
atomic parameters used to derive the CNO abundances are described below.

\subsubsection{Carbon}
\label{sss:C}
Multiple features have been used to derive the C abundances of the Hyades stars
in our sample.  The primary features used for this purpose are two lines of the
C$_2$ Swan system located at 5086.3 and 5135.6 {\AA}, although reliable
measurements could only be made of the latter in the spectra of the giants.  
The C$_2$ features are blends of multiple components of the C$_2$ system, and 
therefore the abundances were derived using the spectrum synthesis method (Figure 
4).  The linelists for these features have been constructed using atomic data from 
the Vienna Atomic Line Database \citep[VALD;][]
{1995A&AS..112..525P,1997BaltA...6..244R,1999A&AS..138..119K,2000BaltA...9..590K}
and C$_2$ molecular data from \citet{1981ApJ...248..228L}.  The transition 
probabilities ($\log gf$) of the C$_2$ and other features have been altered 
slightly from those given by Lambert \& Reis in order to fit the $\lambda 5086$ and 
$\lambda 5135$ C$_2$ features in the Kurucz solar atlas assuming a solar C 
abundance of $A_{\odot}(\mathrm{C}) = \log N_{\odot}(\mathrm{C}) = 8.39$ 
\citep{2005ASPC..336...25A}.
\marginpar{Fig.4}

Carbon abundances of the giants have also been derived using spectral synthesis
of the forbidden [\ion{C}{1}] line at 8727.13 {\AA}.  The $\lambda 8727$ 
[\ion{C}{1}] line arises from an electric quadrapole ($D_2^1 - S_0^1$) 
transition that is strongly coupled through collisions to the \ion{C}{1} ground 
state, which is populated according to Boltzmann statistics.  Its strong 
collisional coupling to the ground state ensures that the [\ion{C}{1}] 
transition also obeys Boltzmann statistics and that the $\lambda 8727$ line 
forms in LTE \citep{1990A&A...237..125S}.  Temperature inhomogeneities due to 
photospheric granulation, the so-called 3D effects, are also not expected to 
greatly affect this line; 3D corrections for the Sun and other solar-type stars 
have been calculated to be $\leq 0.05$ dex \citep{2005ARA&A..43..481A}.  For 
solar-metallicity giants, the correction is of similar magnitude but positive 
(M. Asplund, private communication).  Thus, the $\lambda 8727$ [\ion{C}{1}] 
line is expected to be an accurate abundance indicator for the Hyades giants.  
We adopt the transition probability ($\log gf = -8.136$) for this line 
from \citet{2002ApJ...573L.137A}, and the linelist of the surrounding features 
has been constructed using atomic data from VALD and CN molecular data from 
\citet{1999A&A...342..426G}.  As first identified by
\citet{1967SoPh....2...34L}, the $\lambda 8727$ \ion{C}{1} feature is blended with
a weak \ion{Fe}{1} line at 8727.10 {\AA}.  This \ion{Fe}{1} line has been 
determined to contribute negligibly to the [\ion{C}{1}] line strength in the 
solar spectrum 
\citep[e.g.,][]{1967SoPh....2...34L,1999A&A...342..426G,2002ApJ...573L.137A},
but \citet{2006MNRAS.367.1181B} found that the \ion{Fe}{1} line 
contributes more significantly to the $\lambda 8727$ feature in the spectra of 
stars with $T_{\mathrm{eff}} < 5700 \; \mathrm{K}$ and those at high [Fe/H].  
The Hyades giants ($T_{\mathrm{eff}} \sim 4940 \; \mathrm{K}$, $\mathrm{[Fe/H]} 
= +0.13$) meet both of these criteria, and thus we have included the \ion{Fe}{1}
line in our linelist.  The synthetic fit to the [\ion{C}{1}] line in the 
spectrum of \gtau\ is shown in Figure 5 as an example of our results. 
\marginpar{Fig.5}

In addition to the forbidden [\ion{C}{1}] and the C$_2$ Swan lines, we have 
analyzed ten CH molecular features in the wavelength interval of 4323 -- 4327 
{\AA} in the spectrum of \gtau.  Abundances were derived by fitting a synthetic 
spectrum to each individual line so that ten separate abundance estimates were 
obtained.  Similar to the procedure used for C$_2$, the linelist for the 4325 
{\AA} region has been calibrated by fitting the Kurucz solar atlas assuming a 
solar C abundance of $A_{\odot}(\mathrm{C}) = 8.39$.

The C abundances derived from the various features described above are given in
Table 1.  The mean abundance of the three dwarfs is $A(\mathrm{C}) = 8.54 \pm 
0.03$ (uncertainty in the mean: $\sigma_{\mu} = \sigma_{s.d.} / \sqrt{N-1}$), 
corresponding to a relative abundance of $[\mathrm{C/H}] = +0.15$ when adopting 
the solar abundance of $A_{\odot}(\mathrm{C}) = 8.39$.  The fits to the C$_2$ Swan 
lines in the dwarf spectra are only mildly sensitive to the adopted \teff\ and 
$\log g$ (Table 2), and we estimate the per star uncertainty due to fitting
synthetic spectra to the observed blended features is $\pm 0.05$ dex.  The typical
total per star uncertainty in the derived dwarf C abundances, which is obtained by
adding in quadrature the individual uncertainties due to the adopted parameters 
and synthetic fits, is $\pm 0.07$ dex.  Inasmuch as the standard deviation about 
the mean is an empirical measure of actual scatter introduced by measurement and 
parameter uncertainties, the slight difference in the standard deviation about 
the mean dwarf abundance ($\sigma_{s.d.} = 0.05$) and our estimated total per star
uncertainty ($\pm 0.07$ dex) suggests that, if anything, we slightly overestimate
the expected uncertainties in the {\it relative} dwarf parameters based upon the 
adopted parameter uncertainties listed in Table 1.
\marginpar{Tab.2}  

Our mean dwarf abundance of $[\mathrm{C/H}] = +0.15$ is consistent with previously 
derived values, which range from $[\mathrm{C/H}] = +0.02$ to $[\mathrm{C/H}] = 
+0.18$
 \citep{1978ApJ...223..937T,1990ApJ...351..480F,1999A&A...351..247V,2006AJ....131.1057S}.
All of the previously derived C abundances for Hyades dwarfs are from analyses of 
high-excitation \ion{C}{1} lines; these lines are known to be sensitive to non-LTE 
(NLTE) effects in solar-type stars \citep{2005ARA&A..43..481A} and potentially 
over-excitation effects in late-G and K dwarfs \citep{2004ApJ...602L.117S}.  The 
C$_2$ lines, which arise from electronic transitions, are not expected to be 
influenced by NLTE effects in solar-type stars \citep{2005A&A...431..693A}, and the 
abundances of the Hyades dwarfs presented here should need no NLTE corrections.

The C abundances of the giants, derived from the three different C spectral
features, are in excellent agreement and have a mean value of $A(\mathrm{C}) = 
8.17 \pm 0.01$ (uncertainty in the mean) or $[\mathrm{C/H}] = -0.22$.  The shape 
of the $\lambda 5135$ C$_2$ feature is much less affected by blends in the spectra 
of the giants (Figure 4), and the uncertainty in the synthetic fit is lower than 
that for the dwarfs and is in fact negligible relative to the errors due to the 
adopted \teff.  The typical per star uncertainty in the C$_2$-based abundances of
the giants, dominated by uncertainties in the parameters, is $\pm 0.09$ dex.  
Similar to C$_2$, total per star uncertainties in the [\ion{C}{1}]-based 
abundances are dominated by uncertainties in \teff\ and $\log g$ and have a 
typical value of $\pm 0.11$ dex.  The expected standard deviation about the mean 
giant C abundance is thus $\sim 0.10$ dex, significantly larger than the observed 
value of 0.03 dex; it could be the case that the inferred uncertainty in the 
observed mean abundance is a systematically low fluctuation from the expected 
uncertainty in the mean.  Also, the difference in the observed and expected 
standard deviations about the mean could suggest that the uncertainties in the 
giants' {\it relative} parameters, particularly \teff, are significantly smaller 
than the uncertainties we have adopted (Table 1).

The concordance in the abundances derived from the various C features for each giant 
is not unspectacular given the potential issues involved with deriving abundances 
of evolved stars from the set of spectral lines presented here.  The 
[\ion{C}{1}] feature is blended with a \ion{Fe}{1} line that may be problematic 
in the spectra of metal-rich giants, as described above.  The molecular 
features may also suffer from unidentified or inaccurately modeled blends in 
the spectra of evolved stars, as well as possibly being influenced by 3D 
effects due to their acute temperature sensitivity.  The consistency of the 
abundances suggests that neither blending features, NLTE effects, nor 3D 
effects are a concern for these atomic and molecular lines in the Hyades 
giants.  Our derived mean abundance is in excellent agreement with that of 
\citet{1981ApJ...248..228L}, who found a mean abundance of $[\mathrm{C/H}] = 
-0.20$ ($\sigma_{\mu} = 0.01$) with typical uncertainties of $\pm 0.15$ dex, 
and with the more recent work of \citet{2006A&A...456.1109M}.  In the latter 
study, the authors derived the abundances of 177 local field clump giants, 
including the three Hyades giants included in our study plus the fourth giant 
in the cluster, $\theta^1$ Tau.  These authors find for the Hyades giants a mean 
abundance of $[\mathrm{C/H}] = -0.23$ ($\sigma_{\mu} = 0.03$), with a typical 
uncertainty of $\pm 0.20$ dex.

\subsubsection{Nitrogen}
\label{sss:N}
Adopting the mean C abundances presented above, we use spectral synthesis to 
derive N abundances of HIP 14976, HIP 21099, and the three giants by fitting three 
CN lines at 6703.98, 6706.72, and 6707.53 {\AA} in the ${\lambda}$6707.8 
\ion{Li}{1} region of our high-resolution McDonald spectra.  The CN features are 
too weak, given our resolution and S/N, in the spectrum of the warmer dwarf HIP 
19793 to obtain an accurate N abundance.  We have taken the ${\lambda}6707$ 
\ion{Li}{1} region linelist from \citet{1997AJ....113.1871K} and updated it with 
new atomic data from VALD and Kurucz\footnotemark[7], and refined CN line data from 
\citet{2004AJ....127.1147M}.

\footnotetext[7]{See http://kurucz.harvard.edu}

Slight adjustments to atomic $gf$ values and wavelengths are made to 
simultaneously reproduce as well as possible the very high S/N and resolution 
Kurucz solar flux atlas and  $R\sim60,000$ spectra of the Hyades giants and 
${\alpha}$ Cen A and B obtained by JRK with the McDonald Observatory 2.1-m 
Cassegrain echelle and the University of Hawai'i 2.2-m coude spectrograph, 
respectively, as part of other programs.  Adjustments to the CN $gf$ values from 
\citet{2004AJ....127.1147M} are made by forcing agreement to the solar flux atlas 
alone. The solar-based CN line calibrations assume the input solar C and O 
abundances given in previous and next subsection, respectively and the solar N 
abundance of \citet[][$A_{\odot}(\mathrm{N}) = 7.78$]{2005ASPC..336...25A}.

The linelist is also used to derive Li abundances in the dwarfs and giants from 
the ${\lambda}6707.8$ \ion{Li}{1} resonance feature.  The Li determinations were 
made after C and N abundances are determined for each star as described above 
(we adopt the mean N abundance of HIP 14976 and HIP 21099 for HIP 19793) so that 
these can be used as inputs to the Li synthesis.  We assume no $^6$Li content in 
any of our Hyades stars; the same assumption is made in the solar and ${\alpha}$ 
Cen syntheses used for linelist calibration.  LTE Li abundances are presented in 
Table 1.  Uncertainties in these values are dominated by those in the profile 
fitting and the adopted $T_{\rm eff}$ values, and are ${\pm}0.05$ and ${\pm}0.07$ 
dex for the dwarfs and giants, respectively.

The dwarf N abundances are in good agreement and have a mean value of 
$A(\mathrm{N}) = 7.58 \pm 0.06$ (uncertainty in the mean) and a mean relative 
abundance of $\mathrm{[N/H]} = -0.21$ when adopting the solar abundance of 
$A_{\odot}(\mathrm{N}) = 7.78$.  Nitrogen abundances of Hyades dwarfs in the 
existing literature are scant, having been reported by only two previous studies. 
\citet{1978ApJ...223..937T} derived N abundances of two Hyades F dwarfs from 
near-IR high-excitation \ion{N}{1} lines and found a mean abundance of 
$A(\mathrm{N}) = 8.05$ ($\sigma_{s.d.} = 0.05$) with a total uncertainty for each 
star of $\pm 0.2$ dex.  Adopting the solar abundance $A_{\odot}(\mathrm{N}) = 7.99$ 
of \citet{1978MNRAS.182..249L}, who used the same \ion{N}{1} lines, among others, 
and the same atomic parameters as \citet{1978ApJ...223..937T}, the authors report a 
relative abundance of $[\mathrm{N/H}] = +0.06$.  The other study is that of 
\citet{1998PASJ...50..509T}, who derived a mean abundance of $A(\mathrm{N}) = 
8.34$ ($\sigma_{s.d.} = 0.12$) for nine Hyades F dwarfs from near-IR 
high-excitation \ion{N}{1} lines with typical uncertainties in each measurement of 
0.2 -- 0.3 dex.  These authors adopted a solar of $A_{\odot}(\mathrm{N}) = 8.05$ 
\citep{1989GeCoA..53..197A}, giving a relative abundance of $[\mathrm{N/H}] = 
+0.29$ for the Hyades dwarfs.

It is seen that the three separate estimates of the Hyades dwarf N abundance
are divergent.  \citet{1998PASJ...50..509T} note that their result may be
spurious due to poor data quality and systematic errors in their measured EWs,
and the authors do not assign a high confidence level to their derived value.  
As noted above, both \citet{1978ApJ...223..937T} and \citet{1998PASJ...50..509T} 
analyzed high-excitation \ion{N}{1} lines in F dwarfs which are known to be 
sensitive to NLTE effects \citep{2005ARA&A..43..481A}.  NLTE calculations for the 
$\lambda 8683$ \ion{N}{1} line, a line used by both Tomkin \& Lambert and Takeda et 
al., for 160 F,G, and K dwarfs and subgiants in the solar neighborhood have been
carried out by \citet{2005PASJ...57...65T}; corrections for F dwarfs are on
the order of -0.15 to -0.20 dex.  Applying this correction to the 
\citet{1978ApJ...223..937T} abundance, taking into account a modest NLTE
correction of $\leq 0.05$ dex expected for the solar \ion{N}{1}-based abundance
\citep{2005ARA&A..43..481A}, reduces the difference between their value and 
ours from 0.27 dex to $\sim 0.15$ dex, bringing the two results into 
statistical agreement.  The inferred low N abundance ([N/Fe] $\sim -0.35$) of the 
Hyades dwarfs is at the lower limit of the observed N abundances of metal-rich
dwarfs in the field, which show some scatter around [N/Fe] $\sim 0$ 
\citep{2002A&A...381..982S,2004A&A...421..649I,2005PASJ...57...65T}.  

For the giants there is also good agreement in the derived N abundances.  The
mean abundance is $A(\mathrm{N}) = 7.95 \pm 0.04$ (uncertainty in the mean), 
corresponding to a mean relative abundance of $\mathrm{[N/H]} = +0.17$.  Similar 
to the case for the dwarfs, previous determinations of Hyades giants N abundances 
are limited to two studies.  First, \citet{1981ApJ...248..228L} derived a mean N 
abundance of $\mathrm{[N/H]} = +0.29 \pm 0.04$ from an analysis of all four Hyades 
giants.  Typical uncertainties in the individual abundances are about 0.15 dex.  
Second, the N abundances of \gtau\ and \etau\ were derived by 
\citet{1982A&A...115..145K}, who found $\mathrm{[N/H]} = +0.03$ for the former and 
$\mathrm{[N/H]} = +0.02$ for the later, with uncertainties estimated to be 0.2 -- 
0.3 dex.  Within the combined uncertainties, the three abundance estimates are in 
accord.  The N abundances for both the dwarfs and giants are given in Table 1.

\subsubsection{Oxygen}
\label{sss:O}
Oxygen abundances for the Hyades dwarfs and giants have been derived in our
previous studies \citep{2006AJ....131.1057S,2006ApJ...636..432S}; here we adopt
the [\ion{O}{1}]-based abundances from the former.  Similar to the $\lambda
8727$ [\ion{C}{1}] forbidden line, the $\lambda 6300$ [\ion{O}{1}] line is not
susceptible to NLTE effects, and corrections due to 3D effects are $\leq 0.05$
dex for the Sun \citep{2004A&A...417..751A} and $\leq 0.01$ for 
solar-metallicity giants \citep*{2007A&A...469..687C}.  Consequently, the
[\ion{O}{1}]-based abundances for the Hyades dwarfs and giants should be robust
against NLTE and 3D effects.

The O abundances for the dwarfs and giants are given in Table 1.  The
dwarf mean abundance is $A(\mathrm{O}) = 8.80 \pm 0.04$ (uncertainty in the mean), 
corresponding to a relative abundance of $\mathrm{[O/H]} = +0.11$ when adopting 
the solar abundance ($A_{\odot}(\mathrm{O}) = 8.69$) derived from the 
$\lambda 6300$ [\ion{O}{1}] line as part of our Hyades study.  The mean abundance 
of the stars considered herein does not differ significantly from that of the 
larger sample presented in \citet{2006AJ....131.1057S}, who report a Hyades dwarf 
mean abundance of $\mathrm{[O/H]} = +0.14 \pm 0.02$ based on the analysis of six 
stars.  The mean abundance for the giants is $A(\mathrm{O}) = 8.77 \pm 0.02$ 
(uncertainty in the mean), corresponding to a relative abundance of 
$\mathrm{[O/H]} = +0.08$.  The O abundances of the dwarfs and giants are in good 
agreement.  \citet{1996AJ....112.2650K} also analyzed the $\lambda 6300$ 
[\ion{O}{1}] line in high-resolution spectra of Hyades dwarfs (2) and giants (4), 
and they found mean abundances of $\mathrm{[O/H]} = +0.15 \pm 0.01$ for the former 
and $\mathrm{[O/H]} = -0.08 \pm 0.01$ for the latter.  This 0.23 dex difference is
not corroborated by the results of \citet{2006AJ....131.1057S}, who were able to
show that the then available $\log gf$ value adopted for the \ion{Ni}{1} blend of 
the 6300 {\AA} feature was inaccurate and mainly responsible for the giant-dwarf 
discrepancy reported by \citet{1996AJ....112.2650K}.

\section{DISCUSSION}
\subsection{Observations vs. Models: CNO}
\label{ss:obsVmodels}
Qualitatively, the observed dwarf and giant CNO abundances follow the predicted 
chemical evolution of our stellar model (Figure 3).  Relative to 
the dwarfs, the giants' C is depleted, N enhanced, and O essentially unchanged.  
This abundance pattern is indicative of the CN cycle and first dredge-up mixing. 
The lack of variation in the observed O abundances further indicates that the 
expanding surface convection zone did not extend deep enough to reach material 
processed by the ON cycle.

A more quantitative comparison verifies that the observed N and O abundances are 
in excellent agreement with predictions.  The surface abundances of our model- in 
both mass fraction ($X$) and logarithmic form- before and after the 
first dredge-up are given in Table 3.  The surface \nit\ abundance is predicted to 
increase by a factor of 2.3 after the first dredge-up.  The observed mean N 
abundance (Table 1) of the giants, $A(\mathrm{N}) = 7.95$, is a factor of 2.3 
higher than that of the dwarfs, $A(\mathrm{N}) = 7.58$, in perfect concordance 
with the prediction.  The \oxy\ mass fraction is predicted to change only 
slightly, by less than 3\%, after the first dredge-up.  This small change is not 
manifested in the logarithmic abundances seen in Table 3 because of a similarly 
small dip in the H mass fraction.  Regardless, this minute difference in O 
abundance would be lost in observational uncertainty and is considered negligible. 
Our observed O abundances of the dwarfs and giants, which are statistically 
indistinguishable, agree with the model.
\marginpar{Tab.3}

The consistent observational and computational results do not extend to C.  
Observationally, the Hyades giants have a C abundance that is factor of 2.33 (0.37 
dex) lower than the dwarfs.  The surface \cara\ abundance of the model is depleted 
by a factor of 1.5 (0.19 dex) after the first dredge-up.  The observed C abundance 
of the giants compared to that of the dwarfs is 0.18 dex, or about a factor of 1.5, 
lower than the model prediction.  As a consequence, the sum of C+N+O is not 
constant; the abundance of the giants ($A(\mathrm{C+N+O}) = 8.91$) is only 79\% of 
that for the dwarfs ($A(\mathrm{C+N+O}) = 9.01$).  The model predicts that this sum 
should remain unchanged as a MS dwarf evolves to the RGB giant clump.

The difference in the observed and modeled \cara\ abundances cannot be attributed 
to the random uncertainties in the mean observed abundances, which are 
$\sigma_{\mu} = 0.03$ and 0.01 for the dwarfs and giants, respectively; the 0.18 
dex discrepancy represents a $\sim 6 \sigma_{\mu}$ result.  As discussed in \S 
\ref{sss:C}, the observed standard deviation about the mean abundance
($\sigma_{s.d.} = 0.03$) of the giants is significantly smaller than the expected 
value (0.10 dex) estimated by the total per star measurement and stellar parameter 
uncertainties.  This expected standard deviation, as well as that for the dwarfs 
(0.07 dex), leads to expected uncertainties in the mean abundances of 
$\sigma_{\mu} = 0.05$ and 0.04 for the dwarfs and giants, respectively.  The 
expected uncertainty in the dwarf-giant abundance difference is then 0.064 dex, 
and the observed 0.18 dex discrepancy is still significant at a $\sim 3 
\sigma_{\mu}$ level.

Uncertainties in the mean (actual or expected) are a measure (empirical or 
theoretical) of the internal random uncertainties introduced into the analysis 
from the {\it relative} measurement and parameter (\teff, $\log g$, and $\xi$) 
uncertainties.  The excellent agreement among the dwarf C abundances and among 
the giant C abundances suggest that these random uncertainties are very small, 
especially for the giants since the \teff\ sensitivities of the C$_2$ and
[\ion{C}{1}] abundances are grossly similar but opposite in direction.  If,
however, there are {\it systematic errors} between the dwarf and giant parameter
scales, then abundance differences significantly larger than those expected from
internal random uncertainties may become manifest.  This does not appear to be 
the case here: any systematic change in the dwarf or giant stellar parameters in 
an effort to account for the 0.18 dex discrepancy between the observed C 
abundances and model predictions eliminates existing agreements between 
different C abundance indicators and creates new discrepancies among the 
abundances of other elements.  For instance, the dwarfs \teff\ would have to be 
decreased by 900 K to account for the C abundance discrepancy; aside from being 
highly improbable, such a large \teff\ error would mean that the derived dwarf N 
abundances are overestimated by 1.5 dex.  A change of $+270$ K would be needed 
for the giants \teff\ to raise their C$_2$-based abundance 0.18 dex; this would, 
however, result in a -0.14 dex decrease in the [\ion{C}{1}]-based abundance and 
create a 0.32 dex discordance in the C abundances of each giant derived from 
these two features.  Furthermore, the change in \teff\ would raise the giant N 
abundances by 0.67 dex.  Similarly large ($> 1$ dex) changes in $\log g$ are 
needed to bring the observed C abundances into agreement with the model, and 
similar disagreements among the various elements, particularly O in this case, 
result.

Fitting the synthetic spectra to the observed C$_{2}$ features introduces 
additional error into the derived abundances of the dwarfs.  This can arise from 
both the actual matching of the observed line shapes, which are defined by 
multiple blending lines (Figure 4), and uncertainties in the linelist.  The 
latter is not expected to be a significant contributor to the error in the 
abundances, because the dwarfs in our sample have parameters similar to those 
of the Sun.  The linelist may be more of an issue for the giants; however,
the consistent abundances derived from the CH, C$_2$, and [\ion{C}{1}] lines
suggests otherwise.  As for fitting the $\lambda 5135$ feature in the giant 
spectra, the blending of this line is less severe than for the dwarfs, and the
error due to fitting the observed spectrum is negligible compared to the 
uncertainties due to \teff\ and $\log g$.

As a check of the dwarf C$_2$ results, the C abundance of HIP19793 was derived
from the equivalent widths (EWs) of two \ion{C}{1} lines.  \ion{C}{1} lines in 
the spectra of solar-type dwarfs result from high-excitation transitions, and 
to one degree or another, all form out of LTE \citep{2005ARA&A..43..481A}.  The 
magnitude of the NLTE corrections for these lines is dependent on \teff, $\log 
g$, [Fe/H], and [C/Fe], and for the Sun, the corrections range from -0.03 to 
-0.25 dex (LTE analyses overestimate the abundances), depending on the specific 
line in question \citep{2005A&A...431..693A,2006A&A...458..899F}.  Because of
uncertainty associated with NLTE calculations, \ion{C}{1} lines were not
used to determine the C abundances of the dwarfs and giants.   However, NLTE 
corrections similar to those for the Sun are expected for HIP 19793 because of 
the similarities in their stellar parameters \footnotemark[8].  The 
{\it relative} [C/H] abundance of HIP 19793 as derived from high-excitation 
\ion{C}{1} lines should be a reliable indicator of its C content.  The measured 
EWs of the two \ion{C}{1} lines analyzed and the resulting abundances are 
provided in Table 4.  Atomic data for these transitions were taken from 
\citet{2005A&A...431..693A}.  The mean abundance from the \ion{C}{1} lines is 
[C/H] $= +0.14 \pm 0.02$, in excellent agreement with the C$_2$-based values.  
Combining the C$_2$ and \ion{C}{1} results gives a Hyades dwarf abundance of 
[C/H] $= +0.15 \pm 0.02$ (uncertainty in the mean).
\marginpar{Tab.4}

\footnotetext[8]{The following parameters have been adopted for the Sun :
$T_{\mathrm{eff}} = 5777 \; \mathrm{K}$, $\log g = 4.44$, and $\xi = 1.38$}

The discordance between observed C abundances of clump giants and stellar
evolution model predictions may not to be limited to the Hyades.  
\citet{2006A&A...456.1109M} derived the abundances of numerous elements for a
collection of giants in the Galactic disk- including the four Hyades giants, as 
mentioned in \S \ref{sss:C}- and compared those of C, N, and Na to predictions 
based on STAREVOL stellar evolution models assuming [C/Fe] $= 0$ (Figures 21 -- 
23, therein).  The giants are divided into three metallicity bins defined by 
[Fe/H] $< -0.15$, $-0.15 <$ [Fe/H] $< +0.12$, and [Fe/H] $> +0.12$.  For the 
two more metal-rich bins, the observed C abundances fall below the [C/Fe] $= 0$ 
evolutionary models.  The evolutionary curves fit better the data if the model 
[C/Fe] ratios are shifted by -0.15 and -0.20 dex, respectively, as Mishenina et 
al. show in their Figures 22 and 23; the shifts are motivated by the empirical 
[C/Fe] vs [Fe/H] trend for the giants shown in their Figure 10.

While the initial MS C abundances of the field giants cannot be recovered, the MS
abundances would be expected to be higher than those of the giants, because the 
latter have been diluted by first dredge-up mixing of material processed by 
the CN cycle.  Better indicators of the MS C abundances are those of current MS 
dwarfs at similar metallicities \citep[e.g.,][]{2007AJ....133.2464L}.  C 
abundances of dwarfs have been the focus of numerous high-resolution 
spectroscopic studies focusing on Galactic chemical evolution 
\citep[e.g.,][]{1999A&A...342..426G}.  There is debate as to whether or not 
[C/Fe] ratios increase with decreasing [Fe/H] 
\citep[e.g.,][]{1999A&A...342..426G,2005PASJ...57...65T,2006MNRAS.367.1329R} or 
remain flat \citep{2006MNRAS.367.1181B} at sub-solar metallicities; however the 
behavior near [Fe/H] $= 0$ is consistent: [C/Fe] $\approx 0$.  Down to about [Fe/H] 
$= -0.3$, the results remain consistent with [C/Fe] $\approx +0.10$, and at 
slightly super-solar metallicities, [Fe/H] $= +0.2$, C abundances range from $0 
\geq$ [C/Fe] $\geq -0.10$, depending on the study 
\citep[e.g.,][]{1999A&A...342..426G,2006MNRAS.367.1181B}.  These data, when
compared to the results of \citet{2006A&A...456.1109M}, confirm that C 
abundances of dwarfs are larger than those for giants at a given metallicity. 
Consequently, the [C/Fe] $= -0.15$ and -0.20 stellar evolutionary models that
best fit the data of the two more metal-rich bins of
\citet{2006A&A...456.1109M}, and indeed the [C/Fe] $= 0$ model that fits the 
[Fe/H] $\approx -0.15$ bin, may not be appropriate for those metallicities and
need to be increased by 0.10 to 0.15 dex.  The implication would in that case be 
that the stellar evolution models do not accurately predict the observed C 
abundances of clump giants in the Galactic disk.

A critical diagnostic of stellar evolution is the surface \cara/\carb\ ratio. 
The first reaction of the CN cycle converts \cara\ into $^{13}$N which then
e$^{+}$ decays to \carb, and \carb\ is subsequently converted to \nit\ by 
proton capture.  The \carb(p,$\gamma$)\nit\ reaction lags behind the first 
reaction, so a decrease in \cara\ and an accompanying increase in \carb\ occurs.  
The surface \cara/\carb\ ratio will be attenuated after the first dredge-up 
mixing of core material to the outer layers.  Our stellar model evolves onto 
the MS with a surface \cara/\carb\ ratio of about 90, the solar value, and 
decreases to 23.4 after the first dredge-up.  This prediction matches well the 
observed \cara/\carb\ ratios of the Hyades giants.  \citet{1976ApJ...210..694T} 
analyzed \cara N and \carb N lines near 6300 and 8000 {\AA} in high-resolution 
spectra of the four Hyades giants and found a mean value of 21.0 ($\sigma_{s.d.} 
= 1.8$), and \citet{1989ApJ...347..835G} derived a mean value of 25.8 
($\sigma_{s.d.} = 1.4$) using the same \cara N and \carb N lines near 8000 {\AA}.

The agreement between the observed and predicted \cara/\carb\ ratios of the giants
complicates the interpretation of the discrepant observed and modeled \cara\ 
abundances.  If the \cara\ abundance of the giants has been overdepleted relative 
to standard stellar model predictions, the accurately predicted \cara/\carb\ 
ratios suggest that the overdepletion must extend to \carb, as well.  It is not 
known what mechanism could be responsible for this overdepletion, but in light of 
the agreement between the observed and predicted N and O abundances, it would 
appear that it has to lie outside of the CNO bi-cycle.  Apropos, it is not 
clear how the stellar evolution model can correctly reproduce the observed N, O, 
and \cara/\carb\ results while concurrently failing to do the same for \cara.  
Nonetheless, there is no compelling evidence to suggest that the highly consistent
observed abundances of the dwarfs and giants are erroneous.

\subsection{CAUB Model Sensitivities}
\label{ss:caub}
We have calculated additional stellar models in an attempt to better match the 
observed C abundances of the Hyades dwarfs and giants while at the same time 
conserving the agreement with the N, O, and \cara/\carb\ results.  The model mass,
metallicity, and nuclear reaction rates have all been varied to test their affect 
on the predicted CNO abundance evolution, but none of the changes resolve the
discrepancy with the observed abundances.  The empirically determined masses of the
Hyades giants are well constrained to be 2.2 -- 2.4 M$_{\odot}$ 
\citep{1995AJ....109..780T,2001A&A...367..111D,2006AJ....131.2643A}, so significant
deviations from the adopted mass of our stellar model (2.5 M$_{\odot}$) are not 
expected.  We thus evolved a 2.2 M$_{\odot}$ model and found that the \cara\ 
depletion factor remained unchanged at 1.5, and the \nit\ enhancement factor was 
reduced from 2.3 to 2.1.

The metallicity of the Hyades open cluster has been determined by many groups, and
recent determinations fall in a small range of [Fe/H] $= +0.08$ -- +0.16 
\citep{2003AJ....125.3185P,2005ApJS..159..100T,2006AJ....131.1057S}.  Our stellar
evolution model is characterized by a metallicity of [m/H] $= + 0.10$ (i.e., the
solar composition scaled by +0.10 dex).  Our observations of Hyades dwarfs, however,
indicate that the cluster has a MS composition which includes a slightly more 
enhanced C abundance ([C/H] $= +0.15$) and a low N abundance ([N/H] $= -0.21$).  
Adopting these C and N abundances while keeping the abundances of the other 
elements at [m/H] $= + 0.10$, we have evolved a stellar model with a modified 
metallicity.  The \cara\ depletion factor increased marginally from 1.5 to 1.6; a 
more significant change occurred for the \nit\ enhancement factor, increasing from 
2.3 to 3.8.

In all of the models discussed above, the reaction rates from the NACRE compilation
\citep{1999NuPhA.656....3A} have been used for the CN cycle reactions (please see \S
\ref{s:model}).  The \nit($p,\gamma$)$^{15}$O reaction is the slowest of these, and 
as a result, the preceding \cara($p,\gamma$)$^{13}$N(e$^{+}\nu$)\carb\ and 
\carb($p,\gamma$)\nit\ reactions increase the local \nit\ abundance at the expense 
of \cara.  A new measurement of the reaction rate for the important 
\nit($p,\gamma$)$^{15}$O reaction has been reported by \citet{2005EPJA...25..455I}, 
who find a rate that is a factor of $\sim 2$ lower than that of the NACRE 
compilation.  Incorporating the Imbriani et al. reaction rate into our stellar 
model produces only minor changes in the \cara\ depletion (1.5 to 1.6) and \nit\ 
enhancement (2.3 to 2.2) factors.  We attempted to further increase the \cara\ 
depletion factor by adopting a \cara($p,\gamma$)$^{13}$N rate that is a factor of 3 
higher than the NACRE rate and combining that with the \nit($p,\gamma$)$^{15}$O 
rate of Imbriani et al.  In this model, the \cara\ depletion factor increased to 
1.7, while the \nit\ enhancement factor remained unchanged at 2.3.  In a more 
extreme attempt, we used the previously mentioned increased 
\cara($p,\gamma$)$^{13}$N rate plus the \nit($p,\gamma$)$^{15}$O rate of Imbriani 
et al. decreased by a factor of 5.  This model produced a \cara\ depletion factor 
of 1.9 and a \nit\ enhancement factor of 2.5.  The combination of the modified 
rates results in a small improvement in the discrepancy between the observed and 
predicted level of \cara\ depletion in the Hyades giants, but these rates are well 
beyond the statistical uncertainties of the experimental cross section measurements 
used to determine the reaction rates.  Furthermore, any changes in the reaction 
rates affect only the CNO abundances relative to each other and cannot resolve the
discrepancy between the observed dwarf and giant $A(\mathrm{C+N+O})$ abundances.

Finally, we tested the affect of the mixing length parameter $\alpha$ (and thus the 
efficiency of the convective energy transport) on the \cara\ and \nit\ surface 
abundance evolution of the model.  In all of the models discussed above, $\alpha = 
2$, and new calculations with $\pm 1$ of the adopted value were carried out.  The
changes in $\alpha$ had essentially no effect on the \cara\ depletion and \nit\
enhancement factors.

\subsection{The Evolution of Li: Signs of Extra Mixing?}
\label{ss:liEvolve}
Once our stellar model reaches the MS, most of the initial $^7$Li present in the 
star has been destroyed as a result of convective mixing during its pre-MS 
evolution.  At this stage, any surviving $^7$Li is located near the stellar surface 
at an interior radius of $M_{\mathrm{r}} = 2.45 - 2.50 \; \mathrm{M_{\odot}}$ 
(Figure 1).  For the model with simultaneous nuclear burning and mixing, the star 
arrives on the MS with a $^7$Li surface mass fraction of $1.15 \times 10^{-8}$; for 
the model in which the nuclear burning and mixing are calculated sequentially, the 
$^7$Li surface mass fraction is $1.18 \times 10^{-8}$.  After the first dredge-up 
and before the onset of core He burning, the total amount of $^{7}$Li is diluted to 
surface mass fractions of $1.82 \times 10^{-10}$ in the first model and $1.92 
\times 10^{-10}$ in the second model.  This dilution results from mixing over the 
mass radii of $M_{\mathrm{r}} = 0.34 - 2.50 \; \mathrm{M_{\odot}}$.  Therefore, the 
surface abundance of $^7$Li in both models is diluted by a factor of $\sim 63$ as a 
result of the first dredge-up.  During core He burning, a small amount of $^7$Li is 
destroyed in the inner regions of the star, but the surface abundance is not 
affected by this.  

Using our dwarfs to infer the initial Hyades Li abundance and thus test the 
stellar model predictions for Li evolution in our giants is complicated by the 
pre-MS and MS Li depletion manifest in cool Hyades dwarfs 
\citep{1993ApJ...415..150T}.  Before circumventing this obstacle, we note that our
LTE Li abundances of HIP 19793 and 21099, $A$(Li)$=2.36$ and $1.20$, are in 
outstanding agreement with those derived by \citet{1995ApJ...446..203B}, 2.32 and 
1.29, from reanalysis of the \citep{1993ApJ...415..150T} $\lambda$6707 \ion{Li}{1} 
line strengths using $T_{\rm eff}$ values identical to our own within the 
uncertainties.  The Li abundances of HIP 14976 and HIP 21099 differ by a factor of 
2, a significant result given the ${\pm}0.05$ dex per star uncertainty.  This 
result confirms previous findings of modest but significant star-to-star Li 
scatter at fixed color in cool Hyades dwarfs-- scatter that implicates the action 
of rotationally-induced main-sequence mixing \citep{1993ApJ...415..150T}.

We estimate the initial Hyades Li abundance from the maximum abundances exhibited 
by stars on the warm side of the F-star Li dip and at the G star Li peak in Figure 
3 of \citet{1995ApJ...446..203B}.  Employing the NLTE corrections (for solar 
metallicity) from \citet{1994A&A...288..860C}, typically -0.10 to -0.20 dex, 
suggests an initial Hyades Li abundance of $A$(Li)${\sim}3.15$.  The NLTE 
corrections for the giants are ${\sim}+0.19$ dex.  The NLTE Li abundance 
difference between \gtau\ and the initial cluster value, ${\sim}1.9$ dex, is in 
excellent agreement with the 1.8 dex difference predicted by our model.

However, the Hyades giants evince a spread in Li abundance that is not explained 
by standard stellar models.  The ${\sim}0.35$ dex difference in $A$(Li) between 
\gtau\ and \etau\ is significant given the ${\pm}0.07$ dex per star uncertainty.  
In Figure 6 we overplot the spectra of these 2 giants, showing the marked 
disparity in the strength of their Li features relative to other lines in the 
spectra.  There are several possible explanations of the Li abundance spread in 
the Hyades giants: differences in their evolutionary states, small mass loss 
(${\leq}0.02$ $M_{\odot}$; see Figure 1) in \etau\ prior to the RGB, additional 
mixing during the post-MS not predicted by standard stellar models 
\citep{2004AJ....128.2435B,1990ApJ...352..681C} that depletes Li in \etau , or 
uniform additional mixing for the Hyades giants accompanied by subsequent RGB Li 
production in \gtau\ \citep{2001A&A...374.1017P,2004A&A...424..951P}.
\marginpar{Fig.6}

\citet{2001A&A...367..111D} constructed a high-precision Hertzsprung-Russell 
diagram of the Hyades open cluster using Hipparcos-based secular parallaxes, and 
the authors showed, using a solar-metallicity 631 Myr isochrone of 
\citep{2000A&AS..141..371G}, that the giants are of the same evolutionary state 
and fall squarely on the RGB clump.  Furthermore, the surface gravities of the 
giants, both Hipparcos-based physical gravities \citep{2001A&A...367..111D} and 
spectroscopic gravities \citep{1999A&A...350..859S}, differ by less than 0.10 
dex.  The mass loss hypothesis was originally proposed by 
\citet{1977ApJ...214..124B}, and the small amount required to explain the lower Li 
abundance in \etau\ is very tightly constrained by the steep Li profile of the 
stellar model shown in Figure 1.

If differential additional post-MS mixing is to explain the giant Li abundance 
differences, the remarkable uniformity of the CNO,Na,Mg,Al abundances suggests 
this mixing must be shallow.  Additional valuable constraints can be provided by 
Be and B, which are depleted by proton capture at slightly higher temperatures 
than Li.  Unfortunately, \citet{1977ApJ...214..124B} were only able to derive 
upper limits to Be in Hyades giants, and there is no indication in the work of 
\citet{1998ApJ...499..871D} that potential differences in B abundances in the 
Hyades giants were examined.  New data and updated analyses of Be and B in the 
Hyades giants that could stringently constrain any star-to-star differences and 
the depth of an assumed extra-mixing mechanism to explain such differences would 
have great worth.

Additional theoretical work beyond the scope of that carried out here is needed to 
securely identify the constraints that the uniformity of CNO and Na,Mg,Al (please
see below, \S \ref{ss:NaMgAl}) abundances 
provide on a putative Li production mechanism in \gtau, as well as any 
accompanying nucleosynthetic signatures that may betray such production.  If, for 
example, a hot bottom burning (hbb) process \citep[believed to be responsible for 
Li production in some more highly evolved AGB stars;][]{1992ApJ...392L..71S} is 
at work, then the uniformity of the $^{13}$C/$^{12}$C ratios and $^{14}$N 
abundances in all the giants can only be understood if the $^{13}$C and $^{14}$N 
also produced in a hbb process is negligible or can be transmuted.  In principle, 
the latter can be accomplished via $\alpha$ capture, resulting in both a neutron 
source (from $^{13}$C) and $^{19}$F production (from $^{14}$N).  Thus, $n$-capture 
elements and $^{19}$F abundances (perhaps as well as the $^{26}$Al isotopic 
abundance) may be key diagnostics of such production.  We have compared spectral 
regions containing several features of the light and intermediate mass $s$-process 
elements Sr, Zr, Y, and Ba.  The line strengths in \gtau\ and \etau\ are 
indistinguishable.  Observations of HF features in the near-IR to examine F 
differences in the Hyades giants would be worthwhile future work.

\subsection{The Abundances of Na, Mg, \& Al: Signs of non-Standard Processing?}
\label{ss:NaMgAl}
A final test of the standard stellar evolution model can be made with Na, Mg,
and Al.  The surface abundances of the predominant Na, Mg, and Al isotopes 
($^{23}$Na, $^{24}$Mg, and $^{27}$Al) of intermediate mass MS stars may be 
altered if 1) the NeNa and MgAl cycles are active in the core regions, and 2) 
if the convection zone extends deep enough during the first dredge-up to mix to 
the surface material processed by these cycles.  Our stellar evolution model
does show signs of NeNa and to a lesser extent MgAl cycle processing in the
core; however, it is only the products of the former that get mixed to the
surface during the first dredge-up.  The surface abundance of $^{23}$Na is 
predicted to increase by a factor of 1.38 (0.14 dex) and the surface abundances of 
$^{24}$Mg and $^{27}$Al remain unchanged (Table 3).

We have derived Na, Mg, and Al abundances of the Hyades dwarfs and giants using
an LTE EW analysis.  EWs of a set of presumably unblended 
Na, Mg, and Al lines were determined by fitting Gaussian profiles to the 
spectral features using the one-dimensional spectral analysis routine SPECTRE 
\citep{1987BAAS...19.1129F}.  Only lines that are measurable in both our dwarf 
and giant spectra were utilized, and our final linelist contains three lines 
for each Na, Mg, and Al. Uncertainties in the EWs have been estimated by comparing 
the results of numerous measurements of each line; these uncertainties are 
generally $< 5 \; \mathrm{m \AA}$.  The EW measurements, along with the adopted 
atomic parameters from VALD, are given in Table 5.
\marginpar{Tab.5}

The observed Na, Mg, and Al abundances are given in Table 1.  The abundances are 
given relative to self-consistently derived solar values which negates the effects 
of $gf$ value errors and mitigates NLTE and line blending effects (at least for the
dwarfs).  Solar EWs were measured from a high-quality sky spectrum (S/N 
$\approx 950$) obtained with the Harlan J. Smith 2.7-m telescope and the 2dcoude 
echelle spectrometer at the McDonald Observatory during our 2004 October observing 
run.  Abundance sensitivities to the adopted parameters are given in Table 2 and 
result in typical uncertainties in these abundances of about $\pm 0.10$ dex for 
both the dwarfs and giants.

Similar to CNO, the Na, Mg, and Al abundances show a high level of internal 
consistency for the dwarfs and giants.  The uncertainties in the mean abundances 
are no more than $\sigma_{\mu} = 0.03$ for each element.  The mean [m/H] dwarf 
abundances range from +0.09 -- +0.11 and relative to Fe are essentially solar.  The 
derived abundances of the giants, on the other hand, are found to be markedly 
larger than those of the dwarfs; the mean abundances are [Na/H] $= +0.61 \pm 0.02$, 
[Mg/H] $= +0.55 \pm 0.03$, and [Al/H] $= +0.33 \pm 0.02$ (uncertainties in the
means).  Such large overabundances relative to the dwarfs are in stark contrast to 
the model predictions.  As stated above, Na is the only one of these elements that 
is expected to have its surface abundance affected by first dredge-up mixing, and 
the predicted increase is only 0.14 dex.  Applying this increase to the difference 
in the observed dwarf and giant abundances, a 0.38 dex difference remains.

No existing self-enrichment scenario can explain the large enhancements in Na, 
Mg, and Al abundances of the Hyades giants.  The likely culprit for the highly 
enhanced abundances is unaccounted for NLTE effects.  Unfortunately, targeted NLTE 
calculations for these elements in super-solar metallicity giants have not been 
carried out.  NLTE corrections for Na abundances at [Fe/H] $= 0$ have in general 
been shown to be negative 
\citep{2000ARep...44..790M,2003ChJAA...3..316T,2006A&A...456.1109M}, in line 
with theoretical expectations \citep{2005ARA&A..43..481A}, but positive 
corrections have also been suggested \citep{1999A&A...350..955G}.  In light of
the derived Na overabundances of the Hyades giants compared to the dwarfs, and 
the larger consensus in the literature, negative NLTE corrections for Na seem 
more likely.  

From the large grid of corrections provided by \citet{2003ChJAA...3..316T} we 
can roughly estimate a NLTE Na abundance for the Hyades giants.  The corrections
are dependent on line strength (or metallicity) and vary from line-to-line for a
given star.  This effect seems to be present in our data; the strongest Na line
($\lambda 5683$; Table 5) for the giants consistently results in an abundance that 
is 0.10 - 0.15 dex larger than the other two Na lines (Table 6).  The NLTE models 
of \citet{2003ChJAA...3..316T} are divided into 500 K and 0.1 dex in $\log g$ grid 
steps.  The \teff\ $= 5000$ K model with [m/H] $= 0$ is the most suitable for the 
Hyades giants, but with surface gravities of $\log g \approx 2.6$, they fall in 
between the $\log g = 3.0$ and $\log g = 2.0$ grids.  We adopt the corrections of 
the latter, because the predicted NLTE line strengths better match our observed 
EWs.  Applying the corrections to the individual lines of the Hyades giants 
(Takeda et al. models t50g20m0) and the Sun (Table 5 therein), a mean NLTE 
abundance [Na/H]$_{NLTE} = 0.50 \pm 0.02$ is obtained.  While the line-by-line 
agreement in the relative abundances for each giant is improved, the overall NLTE 
correction amounts to only 0.11 dex, far smaller than the 0.38 dex needed to bring 
the observed Na abundances into agreement with the model.
\marginpar{Tab.6}

The situation for Mg and Al is more uncertain due to the dearth of NLTE
calculations for giants, even at solar metallicities.  Further complicating the
matter is that theoretical considerations generally point to {\it positive} 
corrections for LTE abundances due to large photo-ionization cross-sections of 
the lower excitation levels of Mg and the ground state of Al 
\citep{2005ARA&A..43..481A}, although actual calculations find negative 
corrections for at least some stars and particular transitions 
\citep{2000ARep...44..530S,2007MNRAS.382..553L}.  \citet{2000ARep...44..530S}
calculated NLTE abundances for multiple \ion{Mg}{1} lines, including two
analyzed in this study, and found corrections ranging from -0.05 -- +0.02 dex
for stars with [m/H] $= 0$ and $\log g = 2.5$.  \citet{2007MNRAS.382..553L}
derived Na, Mg, and Al NLTE abundances of field clump giants, and the 
corrections for Mg for stars with parameters similar to the Hyades giants are 
of the same order as those from Shmanskaya et al.  The Al corrections for the
Hyades-type giants are negative, with typical values of about $-0.08$ dex. 
Similar to Na, existing Mg and Al NLTE calculations fall short of the 
differences in the observed abundances of the Hyades dwarfs and giants.

Our stellar evolution model and observations suggest that newly synthesized Na 
may have been mixed from the core regions to the surface of the giants, but as 
\citet{2008AJ....135.2341J} point out, no firm conclusions can be made until
reliable NLTE corrections are available.  If NLTE effects are to account for 
the discrepancies between the dwarf and giant Na, Mg, and Al abundances, 
calculations for super-solar metallicity stars will have to produce corrections 
that are 0.15 -- 0.40 dex larger than those from existing solar metallicity 
calculations.  This may not be unreasonable given the sensitivity of the NLTE 
corrections to line strengths.  Finally, we point out that the observed 
overabundances of Na and Al are common characteristics of open cluster giants. 
\citet{2007AJ....134.1216J} has compiled from the literature Na and Al 
abundances of numerous open cluster giants (Figure 7 therein); the Hyades 
abundances fall squarely among these other data.  As for Mg, many studies find 
near solar abundance ratios for open cluster giants 
\citep[e.g.,][]{2000A&A...360..509H,2004A&A...424..951P}, but large Mg 
overabundances similar to those seen in the Hyades have been observed in other
open clusters \citep[e.g.,][]{2005AJ....130..597Y}.

\section{CONCLUSIONS}
\label{s:cons}
We have utilized high-resolution echelle spectroscopy to derive the light element
abundances of three solar-type dwarfs and three RGB clump giants in the Hyades open 
cluster.   Treating the dwarf abundances as a proxy for the initial composition of 
the giants, the observed abundance patterns have been compared to a stellar 
evolution model calculated with the CAUB stellar evolution code.  The model 
reproduces well the observed N and O abundances, likewise the \cara/\carb\ ratio, 
but it fails to match the observed C abundances.  Whereas the model depletes the MS 
\cara\ abundance by 0.19 dex, the observed mean giant abundance is 0.37 dex lower 
than that of the dwarfs.  A similar offset between observed and modeled C 
abundances of giants in the Galactic disk appears also to exist (see \S 
\ref{ss:obsVmodels}).  Random uncertainties in the mean observed abundances and 
uncertainties related to possible systematic errors between the dwarf and giant 
parameter scales have been ruled out as sources of the 0.18 dex discrepancy in the 
observed and predicted levels of \cara\ depletion in the Hyades giants.  Changes to 
the stellar model parameters fail to significantly improve the disagreement between 
the observations and predictions.

The observed Li abundance of the giant \gtau\ is in excellent concordance with 
the amount of surface dilution predicted by our stellar evolution model.  
However, the $\sim 0.35$ dex spread in the Li abundances of \gtau\ and \etau\ is 
not accounted for by standard stellar models.  The highly consistent CNO, and Na, 
Mg, Al abundances of the giants place stringent constraints on any differential 
mixing or Li production mechanisms that may be proffered to explain the divergent 
abundances.  Whatever the mechanism may be, it is apparent that it must be 
unrelated to the uniformly low \cara\ abundances of the giants.  Be and B 
abundances can be used to further probe possible mixing mechanisms, but 
unfortunately, existing data are unable to provide firm conclusions in this regard 
\citep{1977ApJ...214..124B,1998ApJ...499..871D} .  Additional observations and 
theoretical efforts are needed to make further progress in understanding the Li 
abundances of these giants.

Na, Mg, and Al abundances of the Hyades dwarfs and giants were derived, and all 
three elements are greatly enhanced in the giants relative to the dwarfs.  Our 
standard stellar evolution model predicts that the surface $^{23}$Na abundance will 
show a modest increase ($\sim 0.14$ dex) after the first dredge-up, but it falls 
far short of the observed $+0.52$ dex difference in the dwarf and giant abundances. 
The large Na, Mg, and Al overabundances of the giants cannot be explained by any 
known self-enrichment scenarios, and they are likely due to unaccounted for NLTE 
effects.  Existing NLTE corrections for $^{23}$Na \citep{2003ChJAA...3..316T} 
lower the abundance of the giants by another 0.11 dex, but it too is unable to 
account for the large observed difference.  Current NLTE calculations are limited 
to solar metallicities and below, so targeted calculations of super-solar 
metallicity stars for Na, Mg, and Al are needed.  

Finally, we are left to briefly contemplate the physical implication of the 
disagreement between the observed C abundances of the Hyades giants and the 
stellar evolution model.  Unlike the light element abundance patterns of low-mass 
metal-poor giants brighter than the RGB bump, non-canonical mixing cannot explain
the overdepleted \cara\ abundances of the Hyades giants.  This leaves the
possibility that additional \cara, and possibly \carb, has been processed during
the evolution of the giants.  Given the difference in the observed C+N+O 
abundances of the dwarfs and giants and that the N and O abundances follow the 
model prediction, this depletion, if real, must have occurred via a reaction not
associated with the CNO bi-cycle.  At the core temperatures considered here, there 
is no known reaction outside of the bi-cycle that can destroy \cara.  Thus, our 
observational result may signify that an unknown nucleosynthetic process may be at 
work in metal-rich 2.5 M$_{\odot}$ stars.  Additional fine CNO abundance analyses 
of both MS and evolved stars in open clusters at solar metallicities and above 
will be helpful to further investigate this possibility and continue to test 
standard stellar evolution models.

\acknowledgements
S.C.S. acknowledges support provided by the NOAO Leo Goldberg Fellowship; NOAO 
is operated by the Association of Universities for Research Astronomy, Inc., 
under a cooperative agreement with the National Science Foundation.  J.R.K. 
gratefully acknowledges support for this work by grants AST 00-86576 and AST 
02-39518 to J.R.K. from the National Science Foundation and by a generous grant 
from the Charles Curry Foundation to Clemson University.  We thank Clemson
graduate student Y. Chen for her diligent work on the C$_2$ linelists.  S.C.S. 
also thanks V. Smith and K. Cunha for helpful discussions.

{\it Facilities:} \facility{McD:2.7m (2dcoude)}

%The Bibliography

%Figures
%Stellar Evolution Model- pre-First dredge-up
\begin{figure}
\epsscale{0.75}
\plotone{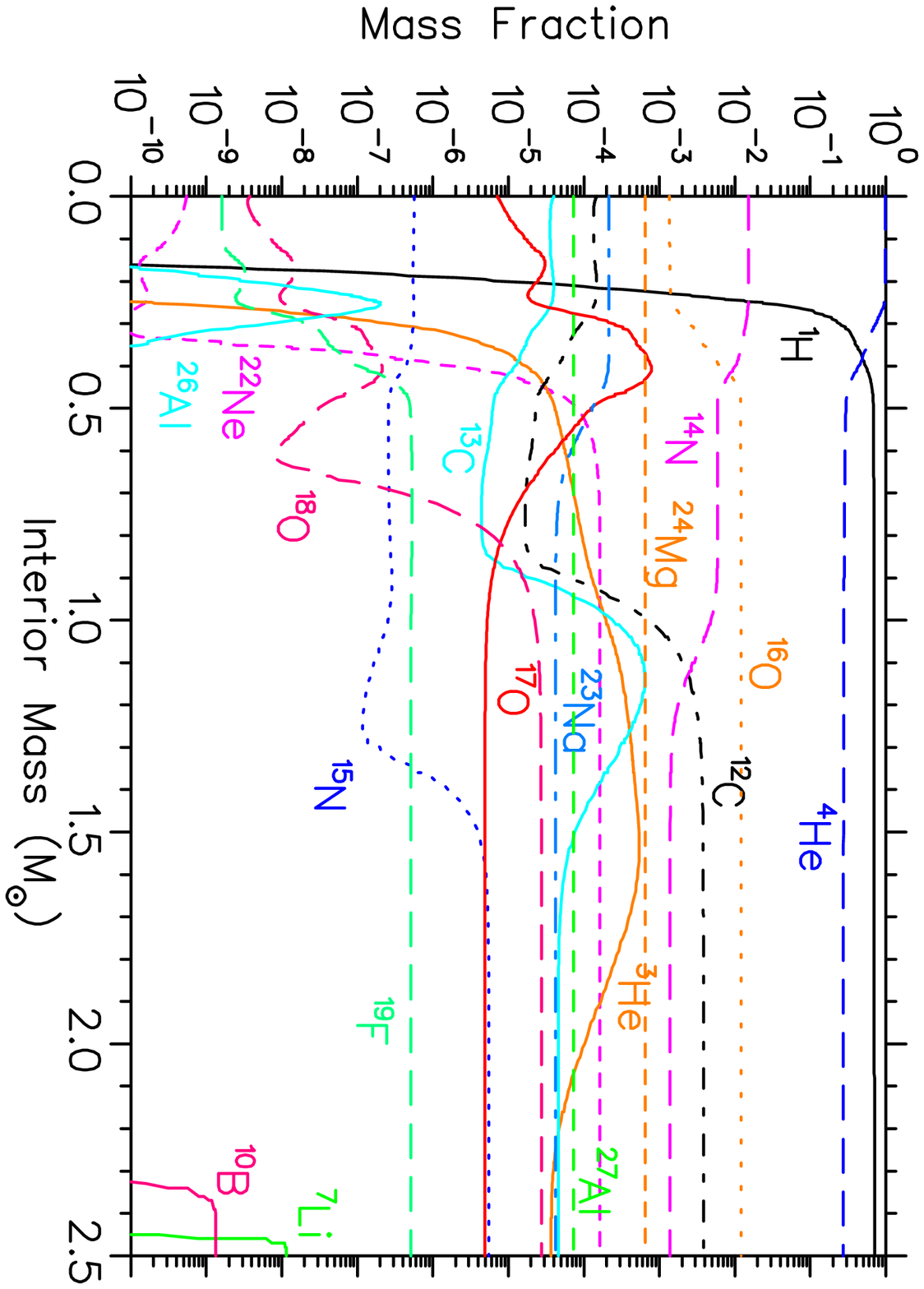}
\caption{Mass fraction vs interior mass of our stellar evolution model before the
first dredge-up.}
\end{figure}

%Stellar Evolution Model- post-First dredge-up
\begin{figure}
\epsscale{0.75}
\plotone{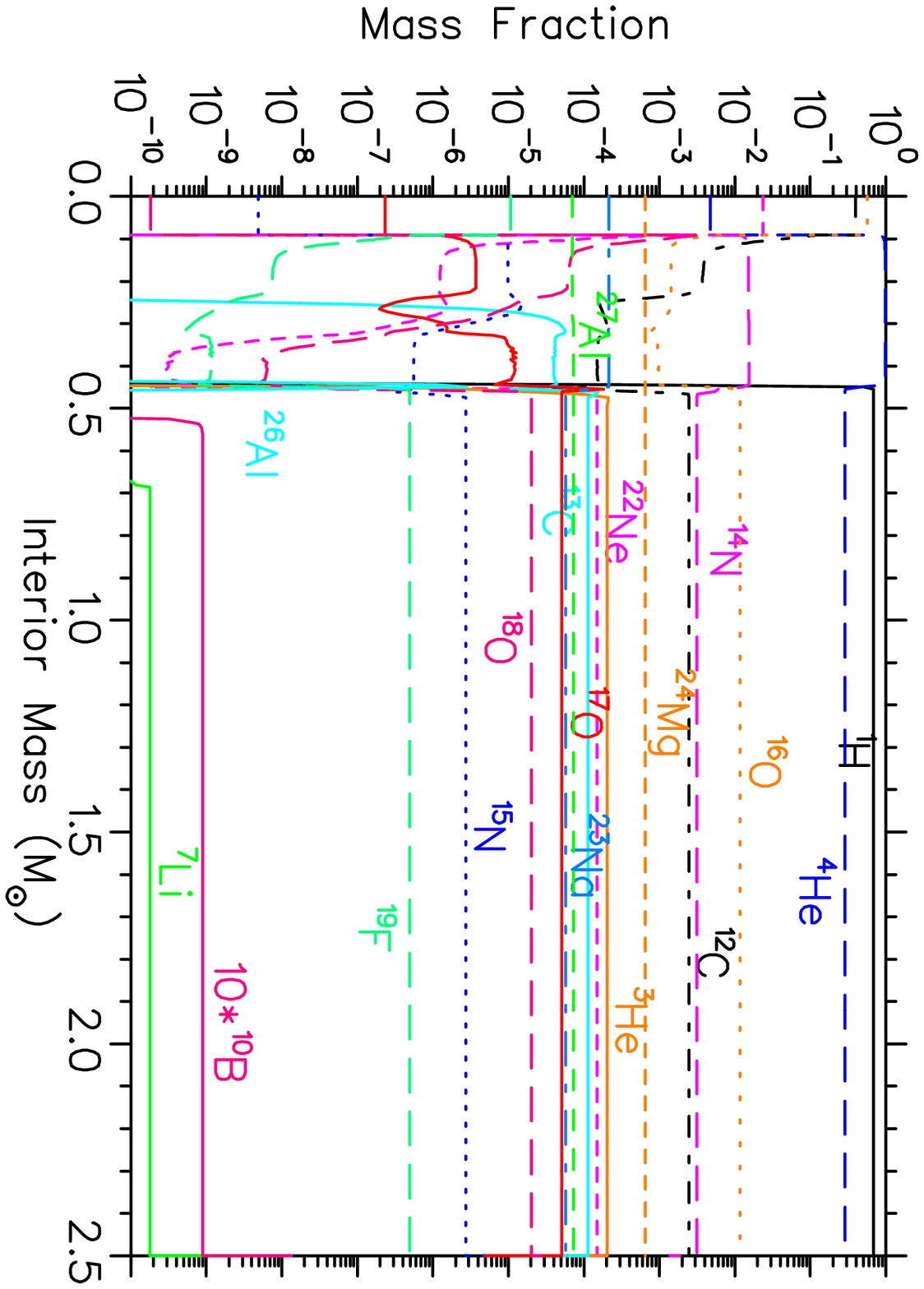}
\caption{Mass fraction vs interior mass of our stellar evolution model after the
first dredge-up.}
\end{figure}

%Surface Composition
\begin{figure}
\plotone{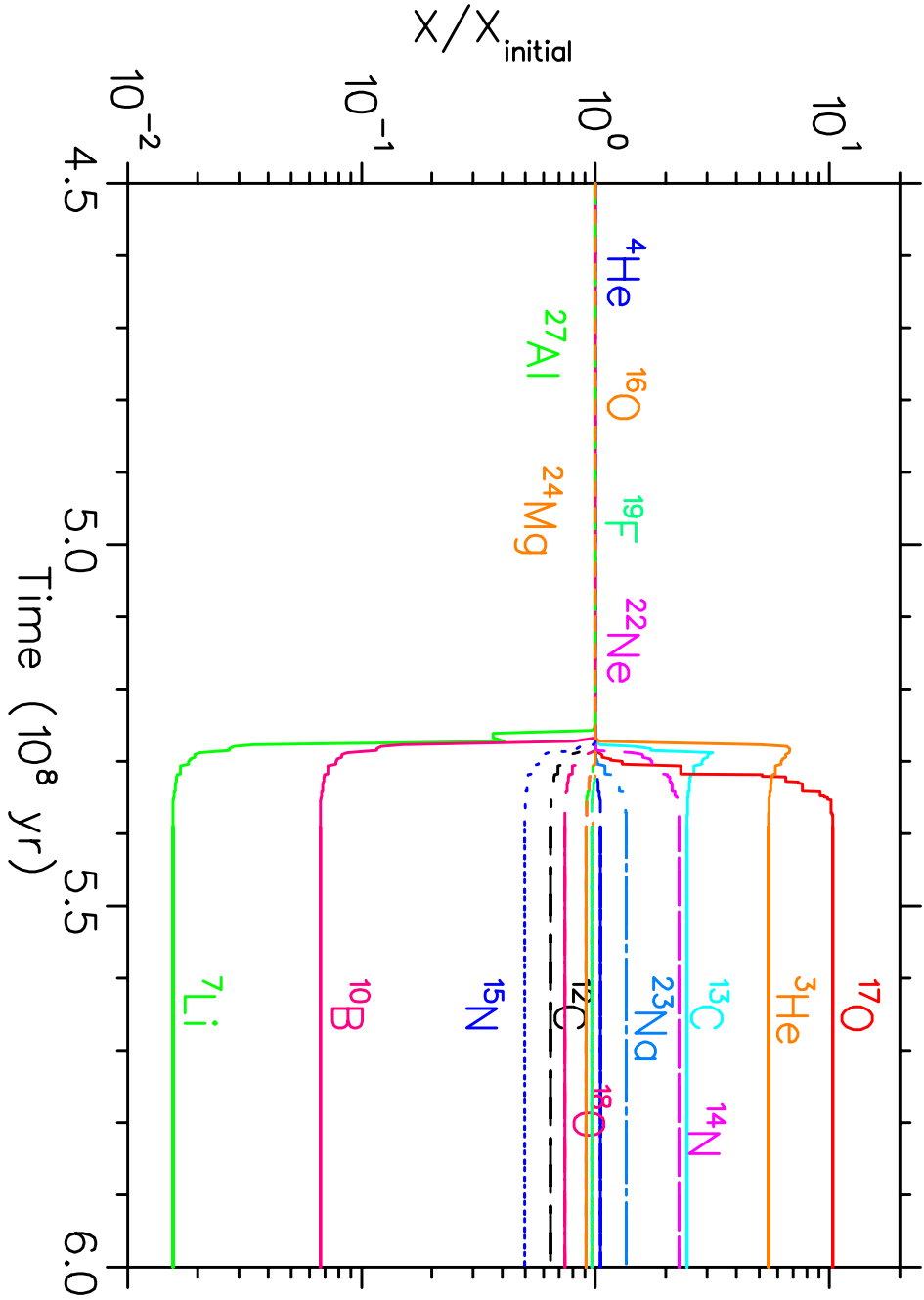}
\caption{The evolution of surface abundances as a function of time. }
\end{figure}

%C_2 Figure
\begin{figure}
\plotone{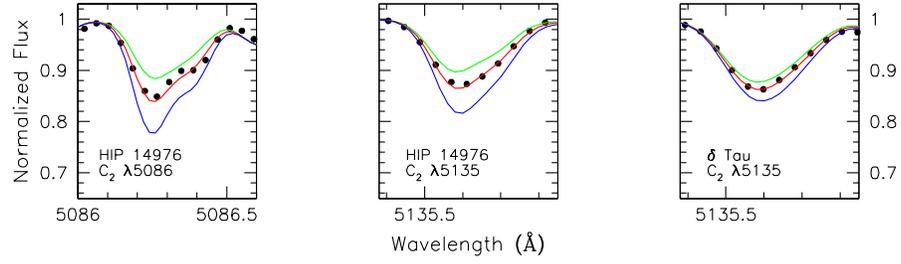}
\caption{Synthetic fits to observed C$_2$ lines for Hyades dwarf HIP 14976 and
giant \dtau.  The red lines represent the best fit abundances for each C$_2$ 
feature and correspond to $A(\mathrm{C}) = 8.54$ and $A(\mathrm{C}) = 8.57$ for 
the $\lambda 5086$ and $\lambda 5135$ features, respectively, for HIP 14976, 
and $A(\mathrm{C}) = 8.15$ for \dtau.  The blue and green lines represent $\pm 0.10$ 
dex of the best fit abundance.}
\end{figure}

%[CI] Figure
\begin{figure}
\plotone{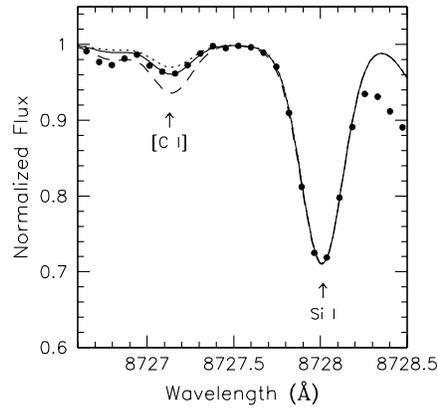}
\caption{The $\lambda 8727$ spectral region of Hyades giant \gtau.  The
[\ion{C}{1}] and strong neighboring \ion{Si}{1} lines are marked.  The best fit
synthetic spectrum is given by the solid line and corresponds to a C abundance of 
$A(\mathrm{C}) = 8.16$.  The broken lines represent syntheses with $\pm 0.10$ dex 
of the best fit abundance.}
\end{figure}

%Li Figure
\begin{figure}
\plotone{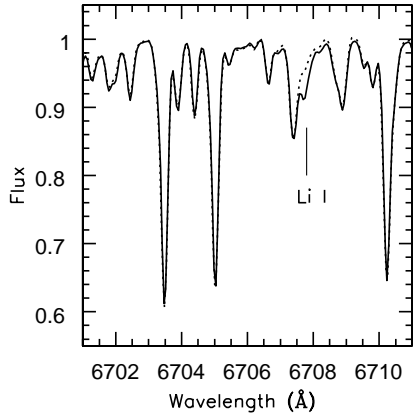}
\caption{${\lambda}6707$ \ion{Li}{1} region spectra of \gtau\ (solid 
line) and \etau\ (dotted line) are overplotted.  The marked difference 
in the strength of the Li feature relative to other lines is apparent.}
\end{figure}

%Tables
\begin{deluxetable}{lcccccccccc}
\tablecolumns{11}
\tablewidth{0pt}
\tablecaption{Hyades Parameters and LTE Abundances}
\tablehead{
     \colhead{}&
     \colhead{}&
     \colhead{HIP 14976}&
     \colhead{HIP 19793}&
     \colhead{HIP 21099}&
     \colhead{}&
     \colhead{$\gamma$ Tau}&
     \colhead{}&
     \colhead{$\delta$ Tau}&
     \colhead{}&
     \colhead{$\epsilon$ Tau}
     }

\startdata
Parameters:                                  &&       &       &       &&      &&       &&		       \\
\phn $T_{\mathrm{eff}}$ (K)\tablenotemark{a} && 5487  & 5722  & 5487  && 4965 && 4938  && 4911        \\
\phn $\log g$ (cgs)\tablenotemark{b}         && 4.54  & 4.49  & 4.54  && 2.63 && 2.69  && 2.57        \\
\phn $\xi$ ($\mathrm{km \; s}^{-1}$)         && 1.24  & 1.34  & 1.24  && 1.32 && 1.40  && 1.47  	       \\
Abundances:                                  &&       &       &       &&      &&       &&		       \\
\phn $\log N(\mathrm{Li})$\dotfill           && 1.51  & 2.36  & 1.20  && 1.06 && 0.93  && 0.72  	       \\
\phn $\log N(\mathrm{C})$\dotfill            &&       &       &       &&      &&       &&		       \\
\phn \phn CH\dotfill                         &&\nodata&\nodata&\nodata&& 8.16 &&\nodata&&\nodata	       \\
\phn \phn C$_2$\dotfill                      && 8.57  & 8.48  & 8.56  && 8.18 && 8.15  && 8.13  	       \\
\phn \phn [\ion{C}{1}]\dotfill               &&\nodata&\nodata&\nodata&& 8.16 && 8.21  && 8.19  	       \\
\phn $\log N(\mathrm{N})$\dotfill            && 7.62  &\nodata& 7.53  && 8.01 && 7.95  && 7.90  	       \\
\phn $\log N(\mathrm{O})$\dotfill            && 8.74  & 8.82  & 8.84  && 8.74 && 8.80  && 8.76  	       \\
%\tablebreak
\phn $\mathrm{[Na/H]}$\dotfill               && 0.06  & 0.13  & 0.07  && 0.63 && 0.58  && 0.61  	       \\
\phn $\mathrm{[Mg/H]}$\dotfill               && 0.09  & 0.09  & 0.12  && 0.57 && 0.50  && 0.57  	       \\
\phn $\mathrm{[Al/H]}$\dotfill               && 0.09  & 0.12  & 0.11  && 0.34 && 0.30  && 0.35  	       \\

\enddata

\tablenotetext{a}{The $1\sigma_{s.d.}$ errors in \teff\ are 55 and 75 K for the
                  dwarfs and giants, respectively.}
\tablenotetext{b}{The $1\sigma_{s.d.}$ errors in $\log g$ are 0.10 and 0.15 dex for the
                  dwarfs and giants, respectively.}

\end{deluxetable}

\begin{deluxetable}{rrrrrrr}
\tablecolumns{7}
\tablewidth{0pt}
\tablecaption{Abundance Sensitivities}
\tablehead{
     \colhead{}&
     \colhead{}&
     \colhead{$\Delta T_{\mathrm{eff}}$}&
     \colhead{}&
     \colhead{$\Delta \log g$}&
     \colhead{}&
     \colhead{$\Delta \xi$}\\
     \colhead{Species}&
     \colhead{}&
     \colhead{($\pm 150 \; \mathrm{K}$)}&
     \colhead{}&
     \colhead{($\pm 0.25 \; \mathrm{dex}$)}&
     \colhead{}&
     \colhead{($\pm 0.30 \; \mathrm{km \; s}^{-1}$)}
     }
     
\startdata
\sidehead{Dwarfs}
C$_2$ && $\pm 0.03$ && $\pm 0.03$ && $0.00$\\
N     && $\pm 0.25$ && $\mp 0.05$ && $0.00$\\
O     && $\pm 0.01$ && $\pm 0.11$ && $0.00$\\
Na    && $\mp 0.09$ && $\pm 0.06$ && $\pm 0.02$\\
Mg    && $\mp 0.08$ && $\pm 0.05$ && $\pm 0.07$\\
Al    && $\mp 0.06$ && $\pm 0.05$ && $\pm 0.02$\\
\sidehead{Giants}
C$_2$        && $^{+0.10}_{-0.18}$ && $^{+0.04}_{-0.01}$ && $0.00$\\
$[$\ion{C}{1}$]$ && $^{-0.08}_{+0.12}$ && $\pm 0.15$ && $0.00$\\
N            && $\pm 0.37$ && $\mp 0.03$ && $0.00$\\
O            && 0.00       && $\pm 0.12$ && $0.00$\\
Na           && $\mp 0.13$ && $\pm 0.05$ && $\pm 0.09$\\
Mg           && $\mp 0.10$ && $\pm 0.04$ && $\pm 0.13$\\
Al           && $\mp 0.08$ && $\pm 0.03$ && $\pm 0.07$\\
\enddata

\end{deluxetable}

\begin{deluxetable}{lccccrr}
\tablecolumns{7}
\tablewidth{0pt}
\tablecaption{Model Surface Abundances}
\tablehead{
     \colhead{}&
     \colhead{}&
     \colhead{$X$}&
     \colhead{}&
     \colhead{$\log N$(m)}&
     \colhead{}&
     \colhead{[m/H]}
     }

\startdata
Dwarf\tablenotemark{a} &&                &&      &&      \\
\phn $^{12}$C  && $3.818 \times 10^{-3}$ && 8.66 && +0.10\\
\phn $^{14}$N  && $1.391 \times 10^{-3}$ && 8.15 && +0.10\\
\phn $^{16}$O  && $1.208 \times 10^{-2}$ && 9.03 && +0.10\\
\phn $^{23}$Na && $4.204 \times 10^{-5}$ && 6.42 && +0.10\\
\phn $^{24}$Mg && $6.481 \times 10^{-4}$ && 7.58 && +0.10\\
\phn $^{27}$Al && $7.300 \times 10^{-5}$ && 6.59 && +0.10\\
Giant\tablenotemark{b} &&                &&      &&      \\
\phn $^{12}$C  && $2.455 \times 10^{-3}$ && 8.47 && -0.08\\
\phn $^{14}$N  && $3.163 \times 10^{-3}$ && 8.52 && +0.47\\
\phn $^{16}$O  && $1.175 \times 10^{-2}$ && 9.03 && +0.10\\
\phn $^{23}$Na && $5.693 \times 10^{-5}$ && 6.56 && +0.24\\
\phn $^{24}$Mg && $6.481 \times 10^{-4}$ && 7.59 && +0.11\\
\phn $^{27}$Al && $7.300 \times 10^{-5}$ && 6.59 && +0.11\\
\enddata

\tablenotetext{a}{Abundances are from model \#20, corresponding to a stellar age
                  of 1.5 kyr.}
\tablenotetext{b}{Abundances are from model \#5890, corresponding to a stellar
                  age of 762 Myr.} 

\end{deluxetable}

\begin{deluxetable}{lcccccc}
\tablecolumns{7}
\tablewidth{0pt}
\tablecaption{HIP 19796 \ion{C}{1} Abundance}
\tablehead{
     \colhead{$\lambda$}&
     \colhead{}&
     \colhead{EW$_{\sun}$}&
     \colhead{EW}&
     \colhead{A(C)}&
     \colhead{}&
     \colhead{}\\
     \colhead{({\AA})}&
     \colhead{}&
     \colhead{(m{\AA})}&
     \colhead{(m{\AA})}&
     \colhead{(dex)}&
     \colhead{}&
     \colhead{[C/H]}
     }

\startdata
5380.34 && 21.7 & 26.0 & 8.68 && +0.16\\
6587.61 && 16.2 & 18.6 & 8.64 && +0.13\\
\enddata

\end{deluxetable}

\begin{deluxetable}{lrcrcrccrrrrrrrrrr}
\tablecolumns{18}
\tablewidth{0pt}
\tabletypesize{\footnotesize}
\rotate
\tablecaption{Na, Mg, \& Al: Spectral Line Data and Equivalent Widths}
\tablehead{
     \colhead{}&
     \colhead{}&
     \colhead{$\lambda$}&     
     \colhead{}&
     \colhead{$\xi$}&
     \colhead{}&     
     \colhead{}&
     \colhead{}&
     \colhead{Sun EW}&
     \colhead{HIP 14976}&     
     \colhead{HIP 19793}&
     \colhead{HIP 21099}&     
     \colhead{}&
     \colhead{$\gamma \; \mathrm{Tau}$}&     
     \colhead{}&
     \colhead{$\delta \; \mathrm{Tau}$}&
     \colhead{}&
     \colhead{$\epsilon \; \mathrm{Tau}$}\\     
     \colhead{Species}&
     \colhead{}&
     \colhead{({\AA})}&     
     \colhead{}&
     \colhead{(eV)}&
     \colhead{}&     
     \colhead{$\log gf$}&
     \colhead{}&
     \colhead{(m{\AA})}&     
     \colhead{(m{\AA})}&
     \colhead{(m{\AA})}&     
     \colhead{(m{\AA})}&
     \colhead{}&
     \colhead{(m{\AA})}&     
     \colhead{}&
     \colhead{(m{\AA})}&
     \colhead{}&
     \colhead{(m{\AA})}
     }    
     
\startdata
\ion{Na}{1}\dotfill && 5682.63 && 2.10 && -0.70 && 104.0 & 131.0 & 122.8 &  129.8 && 169.7 && 171.3 && 171.3\\
                    && 6154.23 && 2.10 && -1.56 &&  38.0 &  55.0 &  47.9 &   56.7 && 104.4 && 102.9 && 108.7\\
		    && 6160.75 && 2.10 && -1.26 &&  58.6 &  76.2 &  69.3 &   78.2 && 120.0 && 120.7 && 125.1\\
\ion{Mg}{1}\dotfill && 4571.10 && 0.00 && -5.69 && 108.5 & 130.3 & 113.9 &  131.8 && 200.7 && 201.0 && 215.5\\
                    && 4730.03 && 4.35 && -2.52 &&  71.2 &  93.4 &  83.8 &   94.8 && 122.0 && 122.6 && 129.0\\
		    && 5711.09 && 4.35 && -1.83 && 105.1 & 123.3 & 113.6 &  127.9 && 157.1 && 156.8 && 161.8\\
\ion{Al}{1}\dotfill && 6698.67 && 3.14 && -1.65 &&  21.0 &  32.5 &  27.0 &   32.9 &&  62.0 &&  61.4 &&  65.5\\
                    && 7835.32 && 4.02 && -0.65 &&  45.0 &  59.2 &  56.0 &   62.8 &&  80.8 &&  78.7 &&  82.8\\
		    && 7836.13 && 4.02 && -0.49 &&  59.3 &  78.5 &  72.8 &   81.3 &&  90.2 &&  90.7 &&  97.8\\

\enddata

\end{deluxetable}

\begin{deluxetable}{lrcrccccccccc}
\tablecolumns{13}
\tablewidth{0pt}
\tabletypesize{\footnotesize}
\rotate
\tablecaption{Na, Mg, \& Al: Line-by-Line Abundances}
\tablehead{
     \colhead{}&
     \colhead{}&
     \colhead{$\lambda$}&
     \colhead{}&
     \colhead{}&
     \colhead{}&
     \colhead{}&
     \colhead{}&
     \colhead{}&
     \colhead{}&
     \colhead{}&
     \colhead{}&
     \colhead{}\\
     \colhead{Species}&
     \colhead{}&
     \colhead{({\AA})}&
     \colhead{}&
     \colhead{Sun\tablenotemark{a}}&
     \colhead{HIP 14976}&     
     \colhead{HIP 19793}&
     \colhead{HIP 21099}&     
     \colhead{$\gamma \; \mathrm{Tau}$}&     
     \colhead{}&
     \colhead{$\delta \; \mathrm{Tau}$}&
     \colhead{}&
     \colhead{$\epsilon \; \mathrm{Tau}$}
     }
     
\startdata
$\mathrm{[Na/H]}$&& 5682.63 && 6.20 & 0.07 & 0.16 & 0.06 & 0.73 && 0.68 && 0.68\\
                 && 6154.23 && 6.28 & 0.07 & 0.12 & 0.10 & 0.61 && 0.54 && 0.59\\
                 && 6160.75 && 6.26 & 0.04 & 0.11 & 0.06 & 0.56 && 0.52 && 0.55\\
		 &&         &&      &      &      &      &      &&      &&     \\
$\mathrm{[Mg/H]}$&& 4571.10 && 7.54 & 0.06 & 0.04 & 0.09 & 0.55 && 0.47 && 0.56\\
                 && 4730.03 && 7.87 & 0.14 & 0.15 & 0.16 & 0.58 && 0.53 && 0.61\\
                 && 5711.09 && 7.61 & 0.06 & 0.08 & 0.11 & 0.57 && 0.50 && 0.54\\
		 &&         &&      &      &      &      &      &&      &&     \\
$\mathrm{[Al/H]}$&& 6698.67 && 6.22 & 0.11 & 0.12 & 0.11 & 0.35 && 0.31 && 0.35\\
                 && 7835.32 && 6.41 & 0.07 & 0.12 & 0.11 & 0.36 && 0.30 && 0.33\\
                 && 7836.13 && 6.43 & 0.10 & 0.13 & 0.12 & 0.31 && 0.28 && 0.36\\

\enddata

\tablenotetext{a}{$A(\mathrm{m})$ abundances are given for the Sun.}

\end{deluxetable}

\end{document}